\documentclass[12pt,review,a4paper]{article}

\usepackage{dcolumn}
\usepackage{bm}
\usepackage{amsmath,amssymb}
\usepackage{subfig}
\usepackage{color}
\usepackage{graphicx}
\usepackage{url}
\pdfoutput=1

\begin{document}

\title{Variational Assimilation for Xenon Dynamical Forecasts in Neutronic 
using Advanced Background Error Covariance Matrix Modelling}

\author{Ang\'elique Pon\c{c}ot$^1$\footnote{angelique.poncot@edf.fr} 
\and Jean-Philippe Argaud$^1$
\and Bertrand Bouriquet$^1$
\and Patrick Erhard$^1$
\and Serge Gratton$^{2,3}$
\and Olivier Thual$^{2,3}$
}

\maketitle
\footnotetext[1]{
Electricit\'e de France,
1 avenue du G\'en\'eral de Gaulle,
F-92141 Clamart Cedex - France
}
\footnotetext[2]{
Sciences de l'Univers au CERFACS, URA CERFACS/CNRS No~1875,\\
42 avenue Gaspard Coriolis,
F-31057 Toulouse Cedex 01 - France
}

\footnotetext[3]{
Universit\' e de Toulouse ; INPT, ENSEEIHT ; 2 rue Camichel , F-31071 Toulouse, France
}


\begin{abstract}

Data assimilation method consists in combining all  available pieces of
information about a  system to obtain optimal estimates of initial states. 
The different sources of information are weighted according to their 
accuracy by the means of error covariance matrices. 
  Our purpose here is to evaluate  the  efficiency 
of variational data assimilation for the xenon induced oscillations
forecasts in nuclear cores. 
In this paper we focus on  the comparison between 3DVAR schemes with 
optimised  background  error covariance matrix $\mathbf{B}$  and a 4DVAR scheme.
Tests were made in  twin experiments  using a  simulation 
code which implements a mono-dimensional coupled model of xenon 
dynamics, thermal,  and thermal-hydraulic processes. 
We enlighten  the very good efficiency of the 4DVAR scheme as well as good
results with the 3DVAR one   using a careful multivariate
modelling of  $\mathbf{B}$ .

{\bf keyword:}  

xenon dynamics, variational data  assimilation,     background error covariance matrix 

\end{abstract}


\section{INTRODUCTION}

In this article we aim to evaluate the efficiency of variational data
assimilation methods known as 3DVAR and 4DVAR methods  in the forecasting of Xenon-135
oscillations.

Xenon-135 is known to be at the origin of axial power oscillations 
of about one day period in pressurised water reactors (PWRs)\cite{Canosa66}.
These oscillations do not change the overall power produced by the nuclear plant 
but they are undesirable from a safety point of view.  
As soon as oscillations are detected, they are damped  using appropriate control rod movements inside the core. 
Detection as well as prediction of xenon induced oscillations are
an important part in the operation of a nuclear power plant. 

No direct measurement of the concentration of xenon in the reactor core is available, and
the simulation of the nonlinear xenon dynamics still 
represents a challenge for real time applications such as system monitoring.
Several models have been proposed for the real time
estimation of xenon concentration. They include flux and iodine-135
dynamics modelling. 
Some of them require an estimation of parameters such as presented in \cite{Onega78-2}. 
This approach is a first step in improving the state estimation but it does not take into account errors in the measurements used
to adjust the model parameters: therefore bad measurements can affect adversely the quality of the computed state.
In addition, it does not allow to correct initial conditions of the coupled flux-iodine-xenon dynamical system. 
Most of these models assume equilibrium concentrations, though
the initial distributions of iodine and xenon have a significant impact on the power transient.

Song and Cho \cite{Song96} determined an analytic initialisation of iodine and xenon of an out-of-equilibrium state which consists in
adding a corrective term with a sinus shape to the 1D equilibrium concentrations. The amplitude
of the sinus is fitted with axial offset power measurements. These measurements are
considered to be perfect, in the same way as the xenon dynamics model is considered exact. This approach is
based on analytical developments which limit the pattern of the added correction and still
does not take into account the errors in the measurements.

Here we aim to improve the Xenon-135 concentration forecast by
finding alternative and more accurate solutions using variational data
assimilation techniques. Such  techniques find their root in earth science and are
used daily in weather forecast. 
%
Data assimilation is nowadays more and more used in the nuclear science community  as well as for the improvement
of the nuclear core activity determination \cite{Bouriquet2010,Bouriquet2011} as for the nuclear accident model parameter determination
\cite{Caccuci10a,Caccuci10b} . Thus it is challenging to apply such a technique to xenon oscillation
forecast even if other trials have been done through genetic algorithm  \cite{Marseguerra03} or optimal
control \cite{Shimazu03}.

Even if the problem we deal with can be seen as very well adapted to the Kalman 
method as it has been shown in \cite{lin94}, we choose to use the variational one instead. This choice has been done as
on long term plan we expect to use  some high resolution models, hopefully together
with their adjoint. Thus it is important to already test variational
techniques that are the only one available when the control space is too large. 

Two data assimilation methods, the 3DVAR and the 4DVAR, will be used here. The
main difference between those methods is that the 4DVAR takes into account the dynamic
of the process. This is a tremendous improvement as it has been proven in weather
forecast. But such a method needs the adjoint that is costly to develop in an industrial context. To
overcome this difficulty a specific model (CIREP1D) compatible with automatic
differentiation has been developed. 

In a first part (section \ref{sec:cirep})  we  describe the model, CIREP1D, a mono-dimensional xenon dynamics model which
includes neutron and thermal-hydraulic processes. Then  we give a brief overview of variational data assimilation methods in
Section \ref{sec:ad}. The setting of the data assimilation method is presented in Section
\ref{sec:ad_neutro}. In Section \ref{sec:B} we propose three background error covariance matrix
modellings (two univariate modellings and a last multivariate modelling). Finally
in Section \ref{sec:res}, we compare the quality of the 3DVAR  estimates  to a
 4DVAR estimate.

\section{THE CIREP1D MODEL}\label{sec:cirep}



Since 3D operational industrial codes are time consuming, we use 
a monodimensional axial xenon/iodine dynamics model coupled with a
 monodimensional neutronic/thermal/thermal-hydraulic model, named CIREP1D.
It simulates axial xenon dynamics  according
to the overall power and the control rod insertion records in a given time window. 
CIREP1D takes a few seconds to simulate a xenon oscillation of a one-week time range but contrary to simpler models,
it gives access to  quantities measured in core:  axial power,  axial xenon, 
axial iodine,  axial flux and   boron  concentration.
 The agreement about axial xenon dynamics between 1D and 3D models is good and has been
studied in detail in  \cite{Poncot08}.



Globally speaking, CIREP1D solves a nonlinear system of  ordinary differential equations given by an operator 
$\mathcal{G}$:
\begin{equation}
\displaystyle \frac{\partial (C_{Xe},C_I)}{\partial t}(z,t) = \mathcal{G}(C_{Xe},C_I)(z,t),
\end{equation}
by using an implicit Euler scheme. $C_{Xe}(z,t)$ and $C_I(z,t)$ represent the axial concentration of xenon and iodine respectively.
Each time step requires a critical boron concentration computation corresponding to the assumption
that neutron and thermal-hydraulic effects may be treated  with stationary coupled equations. 
The xenon dynamics can be initialised 
by either a given xenon and iodine concentrations or by equilibrium concentrations.
Hereafter we give CIREP 1D equations in detail.

\paragraph{{\bf Iodine and xenon equations :}}
Iodine  and xenon balance equations are differential equations on the variables $z,\, t$:
{\small
\begin{equation}\label{eq:cirep:theq}         
\left\{
\begin{array}{rcl}
\displaystyle \frac{\partial C_I}{\partial t}&=&\gamma_I\Sigma_f\Phi -\lambda_I C_I,\\
\mbox{}\\
\displaystyle  \frac{\partial C_{Xe} }{\partial t} &=& \gamma_{X_e} \Sigma_f\Phi
+ \lambda_I C_I  -\left(\lambda_{X_e} +\sigma_{Xe}~\Phi\right)C_{Xe}.
\end{array}
\right.
\end{equation}
}
$\Sigma_f$ is the fission  
cross section of the fuel and $\sigma_{Xe}$ the absorption cross section of xenon-135. These variables depend on $z$ and $t$.
The z-coordinate  is measured from the bottom of the 1D reactor. 
$\gamma_I$ and $\gamma_{Xe}$ are the fractional fission yield of iodine and xenon. Finally, $\lambda_I$ and $\lambda_{Xe}$ are
decay constants of iodine and xenon. 

\paragraph{{\bf Neutronic model:}}
The neutronic flux $\Phi=(\Phi_1,\Phi_2)$ is identified by solving 
two-group diffusion equations. We assume that the time step of flux simulation (a few seconds) is shorter than the xenon 
oscillations. As a consequence, at each time step for the resolution of xenon equation, 
the flux can be computed using the stationary diffusion equations:
{\small
\begin{equation}\label{eq:cirep:eqn}
\left\{
\begin{array}{l}
\displaystyle   -\partial_z D_1 \partial_z \Phi_1 
+ \left[\Sigma_{a1} + \Sigma_r   \right] \Phi_1  = 
\frac{1}{k} \nu \Sigma_f \Phi,\\
\displaystyle \vspace{1.5mm} - \partial_z D_2 \partial_z \Phi_2 + \left[ \Sigma_{a2} + D_2  \right] \Phi_2 - \Sigma_r \Phi_1 = 0,\\
\displaystyle  \mbox{with}\; \nu \Sigma_f \Phi = \nu_1 \Sigma_{f1} \Phi_1 + \nu_2 \Sigma_{f2} \Phi_2. \\
\end{array}
\right.
\end{equation} }
where $\Phi_1$ et $\Phi_2$ are groupwise neutron axial flux distribution, $\Sigma_r$ is the scattering cross section and 
$\Sigma_{ag}$, $D_g$ and  $\nu_g \Sigma_{fg}$ are  the  absorption cross section, 
the  diffusion coefficient, and the  neutron emitted in fission cross section.
All these variables depend on the $z$ variable. The system is closed by using 
albedo boudnary condition.
The balance is obtained by looking for  boron concentration such that
the eigenvalue $k$ is equal to one (this is critical boron concentration computation).
The boron influence  does not appear explicitly in the previous
equations but is linked to  cross section values through the feedback model.

\paragraph{{\bf Feedback model: }}
Thermal effects are not lumped in a power feedback parameter as done in some other cases\cite{Onega78-2}.
The de\-ve\-lo\-ped feedback model is a linear interpolation model relying on  assumption that the cross sections  depend on six quantities:
 fuel irradiation,  xenon concentration $C_{Xe}$, 
 boron concentration $C_B$,  moderator density $\rho_{mod}$,  moderator temperature  $T_{mod}$ and 
fuel temperature $T_f$. Therefore, CIREP1D includes a thermal/thermal-hydraulic model, as described below.

\paragraph{{\bf Thermal-hydraulic model:}}
Since the speed of the water flowing upwards through the reactor is high, we can assume that the thermal-hydraulic problem
is an axial monodimensional problem for the slow transients which are common in the normal operational mode. The moderator temperature $T_{mod}$ is then described by the
following equation:
\begin{equation}\label{eq:cirep:tho}
Q \, \partial_z T_{mod}(z,t)
= \displaystyle \frac{1}{\rho_{mod} c_{mod} } \left[ P_{f}^{lin}(z,t) + P_{mod}^{lin}(z,t) \right],
\end{equation}
where $Q$, $c_{mod}$ and $\rho_{mod}$ respectively represent the volume flow rate, the moderator specific heat capacity  and 
the moderator density. 
Lineic power $P_{f}$ and 
$P_{mod}$ released in both fuel and moderator are computed from the known  two-group flux $\Phi$. 

\paragraph{{\bf Thermal fuel model: }}
Contrary to the thermal-hydraulic model, we employ a radial model for the thermal fuel model. 
Thus, a radial description of the fuel is required.  
We neglect the axial conduction in fuel pin and assume
rotational symmetry of the problem. 
Under these assumptions, the thermal problem can be described
by a monodimensional model in the radial variable $r$:
\begin{equation}\label{eq:cirep:comb}
\displaystyle -
\frac{1}{r}\lambda_{f} \, \partial_r T_{f} (r,z) 
 - \lambda_{f} \, \partial_{r}^2 T_{f} (r,z) = P_{f}(z)/\mathcal{A}.
\end{equation}
The variable $\lambda_{f}$ represents the fuel thermal conductivity and $\mathcal{A}$ corresponds to the pin section. This equation is
coupled with the neutron equation through the lineic power $P_{f}$ and to the thermal-hydraulic problem through the
 boundary condition expressed on edge $\Gamma$:
\begin{equation}
\forall r \in \Gamma, \quad \lambda_f \partial_r T_f(r,z) = h_{tot}(z) \left[ T_{f}(r,z) - T_{mod}(z) \right], 
\end{equation}
where   $h_{tot}$ is  the thermal exchange parameter.
The thermal and thermal-hydraulic parameters $\rho_{mod}$, $c_{mod}$, $\lambda_f$ and  $h_{tot}$ depend on moderator and fuel temperatures. Therefore,
the coupled thermal/thermal-hydraulic problem is nonlinear.

\begin{figure}[!htb]
  \centering
  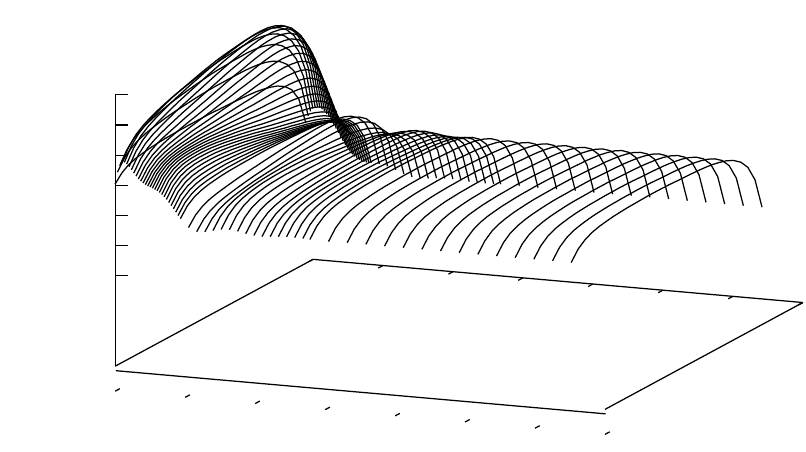
	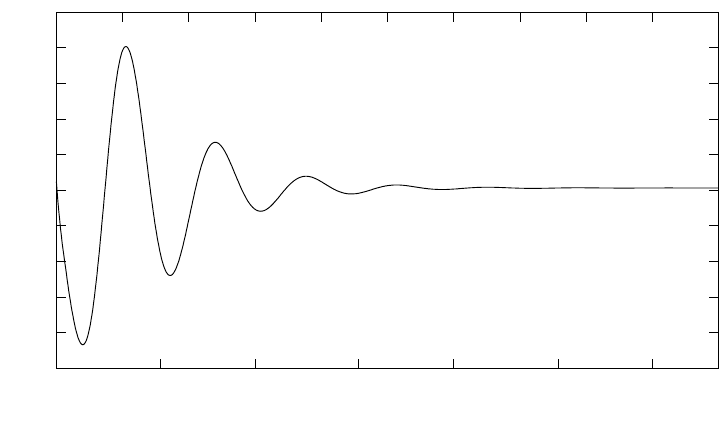
  \caption{ Xenon dynamics simulation on a time range of 200 hours. The map of the first figure represents
 the evolution of the monodimensional xenon vector with respect to the time (x-axis).  The y-coordinate corresponds to the position
in core and the vertical coordinate gives the xenon level. The second plot gives the xenon axial offset with respect to the time, i.e. the power difference between 
the half top and bottom parts of the core.} 
\label{fig:exp}
\end{figure}

\paragraph{{\bf Xenon transient simulation}}
As an example of CIREP1D simulation, we present results from a computation with the following characteristics: time range of 200 hours, 
load following during 30 minutes and, then until the end of simulation, no further rod movement and no power change. 
The  simulated core is in the middle of a burnup cycle and then is moderately irradiated.
Figure \ref{fig:exp} shows a xenon oscillation which disappears without
any external intervention after 100 hours. 
%

\section{A BRIEF OVERVIEW OF DATA ASSIMILATION TECHNIQUES}\label{sec:ad}
Introducing data assimilation method in operational simulation has been an important step to improve 
 forecasts as for example weather in meteorology.
The aim is to provide a satisfactory estimation of the unknown true state of a dynamical system
by combining all pieces of information about the system. 
This information, obtained from measurements (called observations) and simulation,  is
weighted according to its reliability expressed in terms of error covariance
matrices. In practice, the model gives a simulated state called
the background state.
The purpose of data assimilation is to determine a state, called the analysis state, which is
closer to the true state than the one
described solely by the observations or the model. Thus, the analysis state can be used to compute a forecast.

\subsection{Concepts and definitions}
We  now introduce  some concepts and definitions.
A discrete model for the evolution of  physical system  from time $t_i$ to time $t_{i+1}$ is described 
by:
$$ \mathbf{X}(t_{i+1})=\mathcal{M}_{i+1,i}(\mathbf{X}(t_i)),$$
where $\mathbf{X}$ and $\mathcal{M}$ are respectively the model's state vector and its corresponding dynamics operator. 
The dynamics $\mathcal{M}$ of the model evolution is commonly nonlinear. 
We note respectively $\mathbf{M}_{i,j}$ and  $\mathbf{M}^T_{i,j}$ the linear tangent and the adjoint operators with respect to the vector $\mathbf{X}_j$ associated with the dynamics model $\mathcal{M}$ between $t_j$ and $t_i$.
The state vector $\mathbf{X}$ is usually
obtained  by discretisation of physical fields on a grid. 
Its dimension is denoted by $n$. 
The aim  is to evaluate the best estimate of the unknown {\itshape true state}, denoted $\mathbf{X}^t$ 
which is defined by the best possible representation of reality as a state vector at an initial time $t_0$. 
The best estimate that we are looking for in the data assimilation process is called {\itshape analysis} and is denoted by $\mathbf{X}^a$.

The information about the system that can be used to produce the analysis is listed below:
\begin{description}
\item[measurements] in the core gathered into an {\itshape observation vector} $\mathbf{Y}^o$. Its dimension is  $p$.
\item[the observation operator.]
 The key to data analysis is to take advantages of the discrepancy between
observations and state vector. Usually observation vector and state vector are not defined in the same space. 
They can be compared through the use of  a function from model state space to observation space called {\itshape
observation operator} and denoted $\mathcal{H}$. The operator $\mathcal{H}$ can be nonlinear.
\item[an a priori estimate of the true state] before the analysis is carried out. This estimate is called {\itshape background state}
 and is denoted $\mathbf{X}^b$. In most cases,  the analysis problem is under-determined because observations 
are sparse and only indirectly related to the
model variables. The use of this background information helps to make it a well-posed problem and to introduce some physical knowledge.
 Usually, this background state is generated
from the output of a previous analysis.
\item[uncertainties] in the previous data. Background and observation errors are defined by:
$$\epsilon^b=\mathbf{X}^b-\mathbf{X}^t \quad \mbox{and   }\; \epsilon^o_i=\mathbf{Y}^o_i-\mathcal{H}(\mathcal{M}_{i,0}(\mathbf{X}^t)).$$
The covariance matrices of these errors  are denoted by $\mathbf{B}$ and $\mathbf{R}$ respectively.
Error modelling is a difficult task, mostly because  true state $\mathbf{X}^t$ is unknown and 
the  knowledge of the error covariances is  approximative.
But it is a very important step which influences on the quality of the analysis. 
The basic modelling consists in setting up diagonal matrices where the diagonal elements correspond to the variances of the errors
on the background or observation vector. When different fields are involved in the state or observation vector, it is then possible
to choose between an univariate or multivariate modelling. In the former, the errors between the different fields are assumed to be
uncorrelated whereas in the multivariate modelling, the errors are assumed to be correlated. Usually errors between the different kinds of
observation are assumed to be independent and then the covariance matrix $\mathbf{R}$ is diagonal. 
It is more common to assume correlations for the background part but the evaluation of these ones is also difficult.

In the same way, the analysis state $\mathbf{X}^a$ is associated to an analysis error defined by: 
$\epsilon^a=\mathbf{X}^a-\mathbf{X}^t.$ And its covariance matrix denoted by $\mathbf{A}$ is estimated during the assimilation process
or as a postprocessing procedure.

\end{description}



Basically, two families of data assimilation methods exist:  stochastic  and  variational methods.
The most famous stochastic method is probably the Kalman filter. This method is still considered as a reference 
but its application for real data assimilation problems 
is limited to problems of small to medium size due to its huge computational cost involved in matrix computation. Several variants of this method
 have been developed either to reduce its computational cost or to remove the assumption on linearity of the used operators. 
Variational methods are based on the minimisation 
of a cost function \cite{Talagrand97}. These methods, 3DVAR and 4DVAR, that can be adapted  to nonlinear cases  and problems of large size, are mainly
used in operational meteorology and oceanography since the 1990s. Each variational method is equivalent to a stochastic filter method under 
linear assumptions. 
\subsection{Variational methods}
We now give here some elements on variational methods. 
The 4DVAR cost function measures the weighted sum of the square of distances $\mathcal{J}^b$ to background state $\mathbf{X}^b$ and 
$\mathcal{J}^o$ to the observations $\mathbf{Y}^o$ over a time interval $[t_0,t_n]$:
\begin{equation}\label{eq:4D}
 \mathcal{J}_{4DVAR}(\mathbf{X}(t_0))= \mathcal{J}^b(\mathbf{X}(t_0))+ \mathcal{J}_{4DVAR}^o(\mathbf{X}(t_0))
\end{equation}
with
{\footnotesize
$$
\begin{array}{rl}
\displaystyle  \mathcal{J}^b(\mathbf{X}) &=\displaystyle \frac{1}{2}\left[\mathbf{X}-\mathbf{X}^b\right]^T 
\mathbf{B}^{-1}\left[\mathbf{X}-\mathbf{X}^b\right]\\
\displaystyle
\mathcal{J}_{4DVAR}^o(\mathbf{X})&= \displaystyle \frac{1}{2}\sum_{i=0}^{n} \left[\mathbf{Y}_i^{o}-\mathcal{H}(\mathcal{M}_{i,0}(\mathbf{X}))\right]^T\mathbf{R}_i^{-1}\left[\mathbf{Y}_i^{o}-\mathcal{H}(\mathcal{M}_{i,0}(\mathbf{X}))\right]
\end{array}
$$
}
where  weight matrices $\mathbf{B}^{-1}$ and $\mathbf{R}_i^{-1}$ are the inverse of the background and observation error covariance   matrices at time $t_i$.
Minimisation of (\ref{eq:4D}) is done with respect to  initial state $\mathbf{X}(t_0)$. In practice, the starting point of the minimisation algorithm is
taken equal to the background $\mathbf{X}^b$. Evaluations of  gradient of $\mathcal{J}_{4DVAR}$: 
$$
\nabla \mathcal{J}_{4DVAR}\left(\mathbf{X}(t_0)\right)= \nabla \mathcal{J}^b \left(\mathbf{X}(t_0)\right) + \nabla \mathcal{J}_{4DVAR}^o\left(\mathbf{X}(t_0)\right) 
$$
with
{\small
$$
\begin{array}{rl}
\displaystyle  \nabla \mathcal{J}^b\left(\mathbf{X}\right)  = &\mathbf{B}^{-1}\left[\mathbf{X}-\mathbf{X}^b\right] \\
\displaystyle \nabla \mathcal{J}_{4DVAR}^o\left(\mathbf{X}\right) = &-\sum_{i=0}^{n} \mathbf{M}_{i,0}^T \mathbf{H}^T 
\mathbf{R}_i^{-1}\left[\mathbf{Y}_i^{o}-\mathcal{H}(\mathcal{M}_{i,0}(\mathbf{X}))\right],
\end{array}
$$
}
are required by most minimisation methods which implies that the adjoint operator $\mathbf{M}_{i,0}^T$ and
 $\mathbf{H}^T$ have to be evaluated. 3DVAR method is a cheaper alternative  to 4DVAR because 
it does not require the evaluation of the model evolution and its adjoint. The 3DVAR cost function is very close to the 4DVAR one, except for the time
 sum  that disappears:
{\small
$$
\begin{array}{rl}
\displaystyle \mathcal{J}_{3DVAR}(\mathbf{X}(t_0)) =& \mathcal{J}^b(\mathbf{X}(t_0))+ \mathcal{J}_{3DVAR}^o(\mathbf{X}(t_0)) \\
\mbox{with  } \mathcal{J}_{3DVAR}^o(\mathbf{X}) =&
\displaystyle  \frac{1}{2} \left[Y^{o}-\mathcal{H}(\mathbf{X})\right]^T
\mathbf{R}^{-1}\left[Y^{o}-\mathcal{H}(\mathbf{X})\right].
\end{array}
$$
}



In a variational assimilation process, the error covariances of the analysis can be deduced from the  Hessian
 of the cost function $\mathcal{J}_{3DVAR}$ or  $\mathcal{J}_{4DVAR}$\cite{Bouttier99}:
$$
A^{-1}_{4DVAR}=B^{-1}+\sum_{i=0}^n (HM_{i,0})^T_{|X^a_{4DVAR}}R^{-1}(HM_{i,0})_{|X^a_{4DVAR}}.
$$

\subsection{Twin experiment frame}
To validate  assimilation schemes independently of the model, one performs twin experiments.
In twin experiments, the initial true state $\mathbf{X}^t(0)$ is choosen and the true trajectory 
for any time is 
known. It is a simulated state usually obtained by the model used for 
 assimilation. Twin experiments also 
offer  the opportunity to compare the analysis to the true state.
The true state can be used to build  background state, for example by
adding a noise to $\mathbf{X}^t$. It can also be used to build synthetic observations by applying  observation operator $\mathcal{H}$ to $\mathbf{X}^t$
and noise afterwards.

\section{COMPONENTS OF THE ASSIMILATION SYSTEM}\label{sec:ad_neutro}

We develop two variational schemes 3DVAR and 4DVAR, in order to improve the xenon and iodine concentration estimation in core, in twin experiments set-up. 
In this section we describe all the components of the assimilation system except the $\bf{B}$ matrix modelling  that will be discussed in detail in section \ref{sec:B}.

{\bf Model}

The evolution model corresponds to the xenon dynamics model implemented in
CIREP1D.  This model is based on the resolution of the xenon and iodine
mono-dimensional time equations. Each iteration time step requires  a critical
boron concentration computation which includes successive stationary
neutron/thermal/thermal-hydraulic computations. For such a computation, CIREP1D
inputs are:

\begin{itemize}
\item initial and final times $t_0$ and $t_n$ of  transient,
\item initial xenon and iodine concentrations at time $t_0$,
\item transient data: overall power and control rod position variations over the time interval $[t_0;t_n]$.
\end{itemize}

{\bf State vector}

The state vector $\mathbf{X}$ corresponds to  xenon and iodine axial
concentrations discretised on the 30 nodes of the 1D axial spatial mesh used in
CIREP1D. The dimension of $\mathbf{X}$ is then 60. 
The analysis problem is to find a correction $\delta \mathbf{X}$ such that
$\mathbf{X}^a=\mathbf{X}^b+\delta \mathbf{X}$ is as close as possible to
$\mathbf{X}^t$. This correction is searched in the same space as the state
vector one. Thus, the minimisation problem has dimension~60.

{\bf The observation operator $\mathcal{H}$ }

In the present use of data assimilation the  observation operator $\mathcal{H}$
is given by the model itself. It is non-linear
and roughly corresponds  to a critical boron calculation with
CIREP1D.   Therefore, it depends on xenon  but not on iodine as no time
evolution are done in such a calculation. Since the 3DVAR scheme does not involve any evolution model, it cannot
control iodine concentration.
Another important characteristic of this scheme, 
 is the quasi-equivalence in  computational cost of
 evaluation of the model $\mathcal{M}$ and  observation operator $\mathcal{H}$.


{\bf Observations}

In this framework, observations used further in the analysis process are not
coming from real core measurements.  They come from numerical simulations with
CIREP1D, in a twin experiment framework  as described in the section \ref{sec:ad}.
 The scheme is the following:

\begin{enumerate}
\item We compute xenon dynamics initialised by equilibrium concentrations, 
in a time range of one hour for example. The concentrations obtained
after this hour are defined as the real state $\mathbf{X}^t$ 
at the initial time $t_0$ of the future analysis process.
\item We do a reference simulation with CIREP1D to  make the real state $\mathbf{X}^t$ evolve from $t_0$ to $t_n$:
$$\mathbf{X}^t(t_i)=\mathcal{M}_{i,0}(\mathbf{X}(t_0)).$$
\item Observations over  time range are obtained by introducing a measurement noise $\epsilon$ on real data:
$$ \mathbf{Y}_i^o=\mathcal{H}(\mathbf{X}^t(t_i))+\epsilon.$$
\end{enumerate}

The observation vectors $\mathbf{Y}^{obs}_i$ at different observation times $t_i$
are composed of 3 different measurements:  6 integrated powers  over several
cells (index $P$), 1 power axial offset data (index $AO$) and 1 boron
concentration data (index $C_B$). The associated error standard deviations are
denoted respectively by $\sigma_{Rp}$,  $\sigma_{R_{AO}}$ and $\sigma_{RC_b}$.
The observation vector has dimension 8. 

{\bf Error covariance matrices}

To build the  observation  error covariance matrix
$\mathbf{R}$,  we assume that  measurement errors are 
 Gaussian, that they are not correlated in space, and that they does not depend on time.
In this case $R_i=R$ and R is diagonal and given by:
{\small 
\begin{equation}\label{eq:sch:r}
\forall \, 0 \leq i < n_{obs}, \; \mathbf{R} =  \left( \begin{array}{ccccc} \sigma_{Rp}^2  &  & & &0 \\
 & \ddots & & &\\   &  & \sigma_{Rp}^2 & &  \\   & & & \sigma_{R_{AO}}^2 &\\
 0 & & && \sigma_{RC_b}^2   \end{array}  \right).
\end{equation}
}

Errors
introduced in measurements are set to 10\%, 5\% and 1\% of the current value
respectively for axial power ($\sigma_{Rp}$), power axial offset 
$(\sigma_{R_{AO}}) $ and boron concentration ($\sigma_{RC_b}$) measurements. Those values correspond to
the typical knowledge we got on those measurements considering both their intrinsic
error and the representativity error.

{\bf Minimisation}

Finally, to solve the non-linear minimisation problem, we use the quasi-Newton
method LBFGS \cite{Liu89}. This method requires the computation of the gradient
of $\mathcal{J}$ which is done using the adjoint  of the xenon dynamics model. 
In our case, the adjoint is obtained by automatic differentiation of CIREP1D
using TAPENADE software \cite{tapenade}. Both $\mathcal{J}$ and $\nabla
\mathcal{J}$ are computed in the framework of the PALM assimilation coupler
\cite{palm11,Lagarde01}.

\section{BACKGROUND ERROR COVARIANCE MATRIX MODELLING}\label{sec:B}

The $\mathbf{B}$ matrix is one of the most important point of the data assimilation, in
particular for the 3DVAR method.
 For the 4DVAR method, the matrix is less crucial since the model itself  contributes to
spread the information. 
Thus in order to
 have a reliable comparison between both methods a careful study  of  $\mathbf{B}$ is presented here.\\

We recall that the state vector of size 60 is composed by the values of the
physical fields xenon and  iodine at the 30 mesh nodes (finite differences
discretisation), the mesh node numbering starting from the bottom  of the core.
The vector ${\epsilon}^b$ of size $60$  
represent the background error made on each of the 30 mesh nodes,
which is assumed Gaussian. The error covariance matrix $\mathbf{B}$ is defined
by:

$$ \mathbf{B}= E\left[ \left( \mathbf{\epsilon}^b -
E\left[\mathbf{\epsilon}^b\right]\right)\left( \mathbf{\epsilon}^b -
E\left[\mathbf{\epsilon}^b\right]\right)^T\right]. $$

We study three different types of modelling for the background  covariance
matrix $\mathbf{B}$: an elementary modelling where $\mathbf{B}$ is diagonal, a
univariate modelling where $\mathbf{B}$ is block diagonal and at last a
multivariate modelling.

\subsection{Settings used for the modelling}

In what follows, we need to use the evolution model $\mathcal{M}$ and then to
set up a simulated case. We are working in the twin experiment framework. The
simulated  case is a regular transient where the produced power is close to the
nominal power and control rods are partially inserted in core. The state of the
core corresponds to the end of a fuel cycle. 
The true and background trajectories
$\mathbf{X}^t(t)$ and $\mathbf{X}^b(t)$ are computed with non equilibrium
initial states issued from close but not identical previous calculations.  

We use standard deviations $\sigma_{Xe}$ and  $\sigma_I$ close to 3\% for the
initial  $\mathbf{B}$ matrix. Those values can evolve respect to the treatment
we do on this initial matrix. Those values come from comparison  between CIREP1D
and other models. Moreover they are in the typical range of values used in~\cite{Argaud09d,Bouriquet2010,Bouriquet2011}.

\subsection{Elementary modelling}

As a first step to build the $\mathbf{B}$ matrix, we omit correlation in space
and between species consider the diagonal matrix given by:

$$
\mathbf{B}_d = \left(  \begin{array}{cccc} \sigma_{Xe}^2 & 0 & 0 & 0 \\ 0 & \ddots & 0& 0 \\ 0 & 0 & \sigma_I^2 & 0 \\  0 & 0 & 0 & \ddots \end{array}  \right).
$$

\subsection{Univariate modelling}\label{sec:um}

Before developing a multivariate modelling, we propose to take into account
spatial correlations for xenon and iodine. We are looking for a block diagonal 
matrix:

$$ \mathbf{B}_u = \left( \begin{array}{cc} \mathbf{B}_{Xe} &  0  \\ 0 &
\mathbf{B}_{I} \end{array} \right), $$

where the block diagonals are given by:

$$
\begin{array}{l}
\mathbf{B}_{Xe} = \mathbf{B}_{d,Xe} \mathbf{\Gamma}\mathbf{B}_{d,Xe},\\
\mathbf{B}_I = \mathbf{B}_{d,I} \mathbf{\Gamma}\mathbf{B}_{d,I}.\\
\end{array}
$$

The matrices  $\mathbf{B}_{d,Xe}$ and $\mathbf{B}_{d,I}$ correspond to the
sub-matrices extracted from $\mathbf{B}_d$. The matrix $\Gamma$ is built thanks
to the Balgovind correlation\cite{Gaspari99} between two nodes of the spatial
mesh $z_i$ and $z_j$ numbered from $1$ to $30$, which reads:  $$\Gamma(z_i,z_j)
= ( 1 +  |z_i-z_j| /L) \exp( -|z_i-z_j| /L)$$ where the parameter $L$
corresponds to the correlation scale to set up.

This modelling assures
the definite positivity of $\mathbf{B}_u$. The choice of $L$ has consequences
in: 

\begin{itemize}
\item the structure of $\mathbf{B}_u$ (decay property of the matrix elements away from the main diagonal), 
\item the conditioning of $\mathbf{B}_u$ (the larger $L$ is, the worse the conditioning is),
\item and the quality of the analysis. 
\end{itemize}
In practice, the choice of $L=4$ for both species  gives satisfactory results \cite{Poncot08}.

\subsection{Multivariate modelling}

 We propose to build correlations thanks to the evolution model
$\mathcal{M}_{i,0}$ between times $t_0$ and $t_i$.  This method is very close
to what is done in Kalman filter. However we do no consider each step of
evolution.

If we consider a small
initial perturbation $\epsilon$ on the  state $\mathbf{X}$ at time $t_0$, one
can write:

$$
\mathcal{M}_{i,0}(\mathbf{X}(t_0)+\epsilon(t_0)) \approx \mathcal{M}_{i,0}(\mathbf{X}(t_0))+ \mathbf{M}_{i,0|\mathbf{X}}.
\epsilon(t_0),
$$

where $\mathbf{M}_{i,0|\mathbf{X}}$ represents the tangent linear of $\mathcal{M}_{i,0}$ with respect to the vector 
$\mathbf{X}$. Then, if we set  
$\epsilon(t_i)=\mathcal{M}_{i,0}(\mathbf{X}(t_0)+\epsilon(t_0)) -\mathcal{M}_{i,0}(\mathbf{X}(t_0))$, the last relation can 
be expressed in terms of the error as follows:

$$
\epsilon(t_i)   \approx \mathbf{M}_{i,0|\mathbf{X}}.\epsilon(t_0).
$$

Then we get an approximation of the covariance matrix at $t_i$ as a function of the covariance matrix at $t_0$:
\begin{equation}\label{eq:cov}
cov(\epsilon(t_i),\epsilon(t_i))\approx\mathbf{M}_{i,0|\mathbf{X}} cov(\epsilon(t_0),\epsilon(t_0))\mathbf{M}_{i,0|\mathbf{X}}^T.
\end{equation}
It is  noted that it can be difficult to get the tangent linear operator $\mathbf{M}$ for an industrial code, since it usually 
requires to be written at the same time as the direct code.
Thanks to Eq. (\ref{eq:cov}) we can model correlations between xenon and iodine by multiplying an univariate matrix $\mathbf{B}_u$
by $\mathbf{M}_{i,0|\mathbf{X}}$ and $\mathbf{M}_{i,0|\mathbf{X}}^T$:

$$
\mathbf{B}_i=\mathbf{M}_{i,0|\mathbf{X}} \mathbf{B}_u \mathbf{M}_{i,0|\mathbf{X}}^T.
$$

In what follows, we compare the univariate matrices $\mathbf{B}_d$, $\mathbf{B}_u$  to the multivariate matrices $\mathbf{B}_3$,  $\mathbf{B}_{12}$
and $\mathbf{B}_{24}$.

\subsection{Overview of the estimated covariance matrices}

\begin{figure}[htbp]
\begin{center}
\subfloat[$\mathbf{B}_{d}$]{\label{fig:varBd}{\includegraphics[width=0.5\columnwidth]{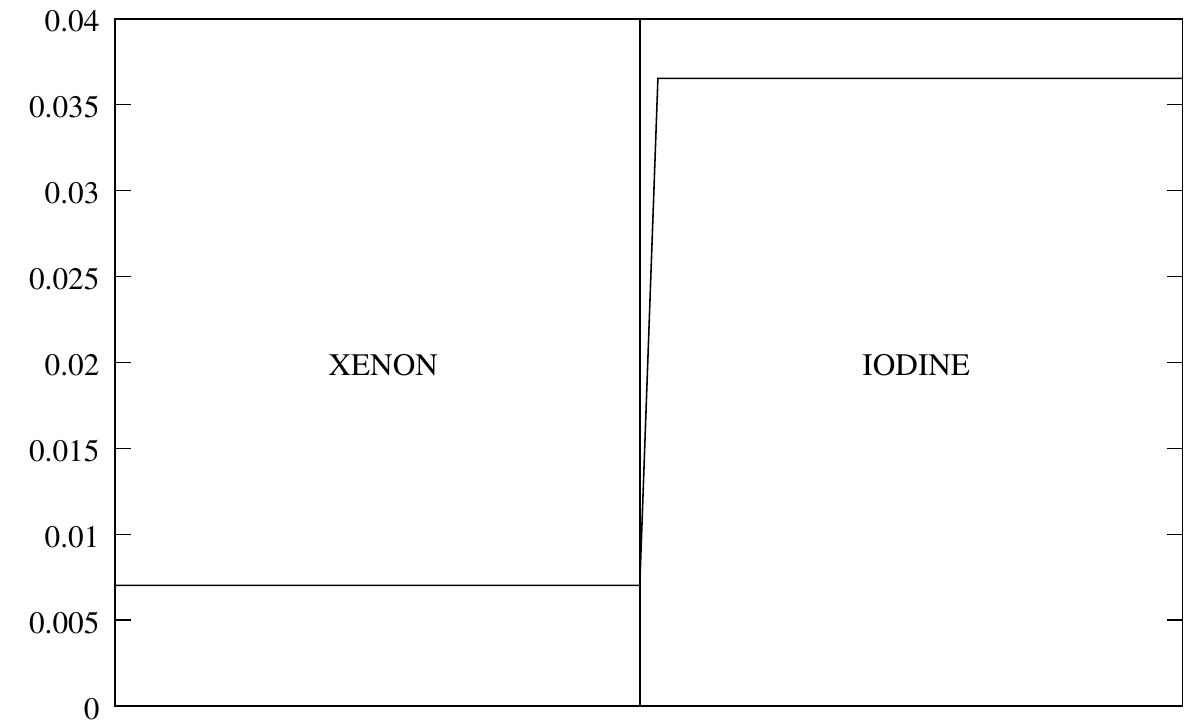}}}\subfloat[$\mathbf{B}_{u}$]{\label{fig:varBu}{\includegraphics[width=0.5\columnwidth]{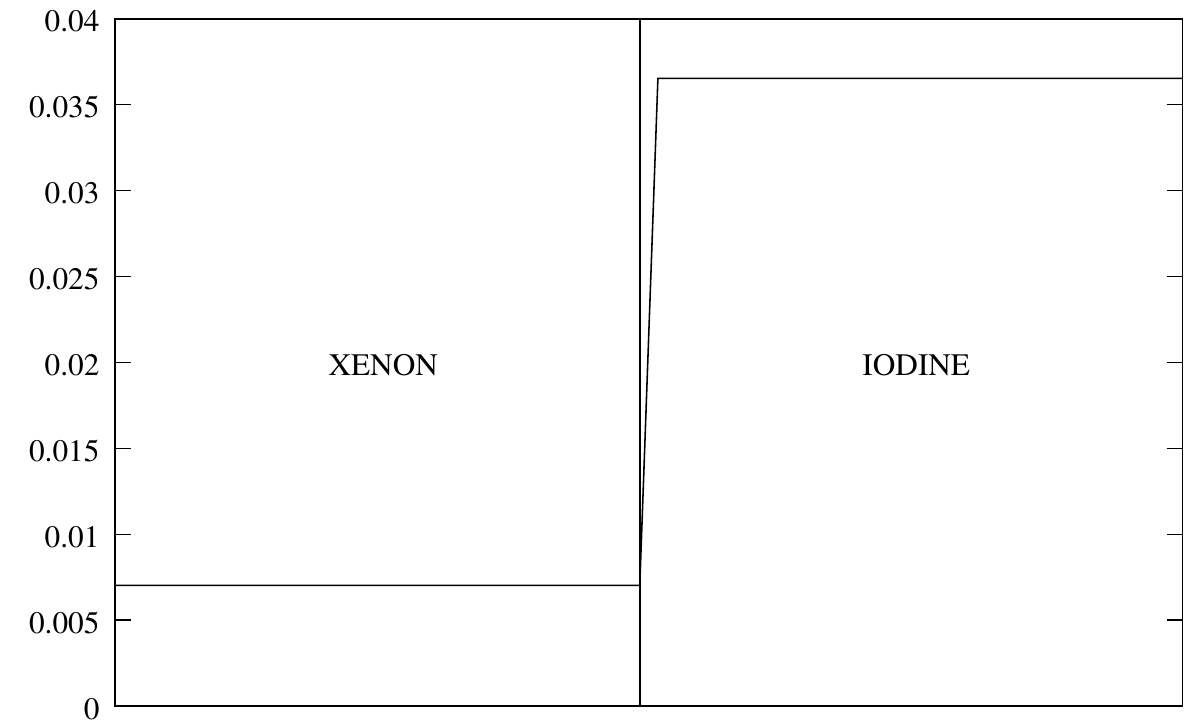}}}\\
\subfloat[$\mathbf{B}_{3}$]{\label{fig:varmB3}{\includegraphics[width=0.5\columnwidth]{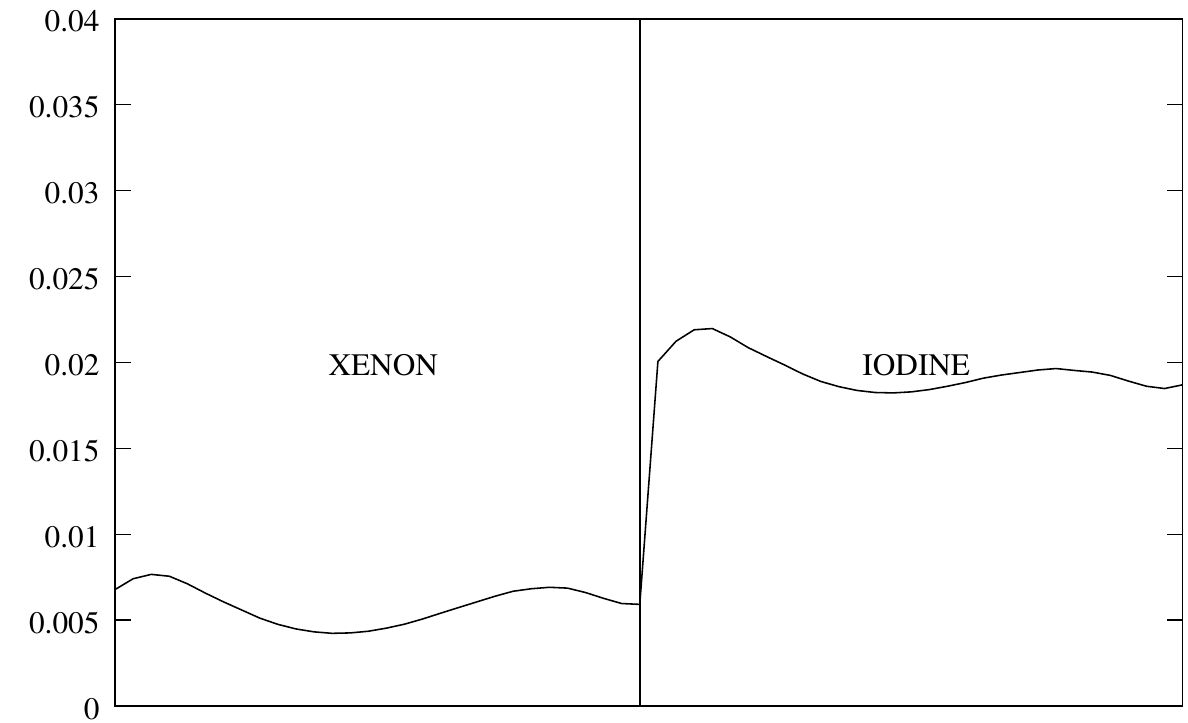}}}\subfloat[$\mathbf{B}_{12}$]{\label{fig:varmB12}{\includegraphics[width=0.5\columnwidth]{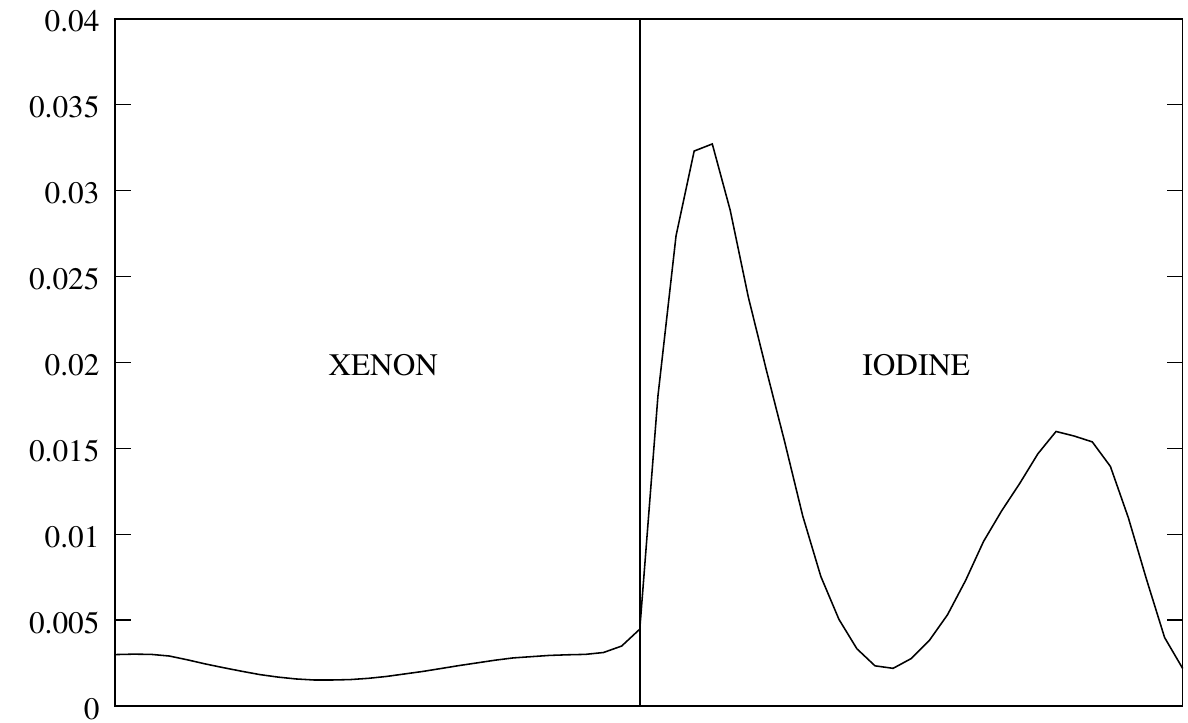}}}\\
\subfloat[$\mathbf{B}_{24}$]{\label{fig:varmB24}{\includegraphics[width=0.5\columnwidth]{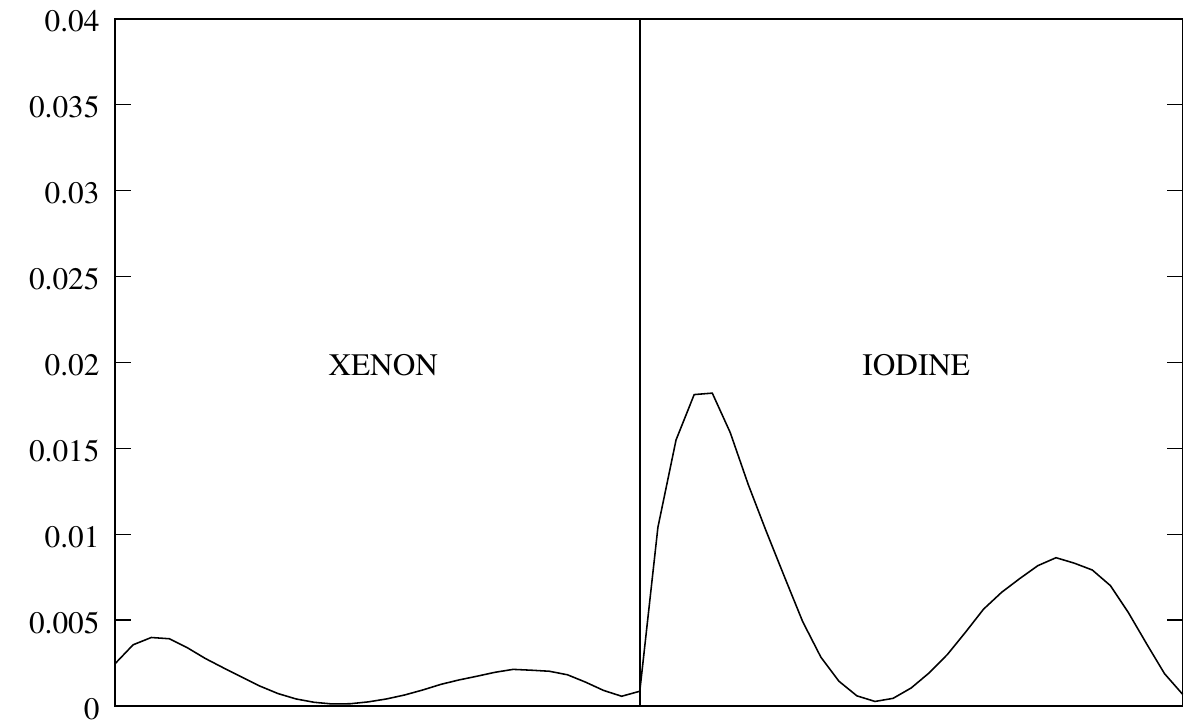}}}
\end{center}
\caption{Diagonal terms of $\mathbf{B}$ matrices for univariate and multivariate modelling, that is to say background error variances with respect to the node
of the spatial mesh.}
\label{fig:varB}
\end{figure}

Figure \ref{fig:varB}  shows the diagonals of the various $\mathbf{B}$ matrices. These diagonals correspond to the error 
variance of the background state vector for each node of the spatial mesh (x-axis in the Figure \ref{fig:varB}). We may notice:
\begin{itemize}
\item In the diagonal and univariate modellings (Figures \ref{fig:varBd} and \ref{fig:varBu} ), 
the variance is the same for all the spatial discretisation points, though it is known that the lower and upper regions of 
the core are quite inaccurately modelled. The iodine error variance is  bigger than the xenon one but it is due to the fact that iodine
 concentration in core is bigger than the xenon concentration. In terms of relative errors, they are similar.
\item For multivariate modellings (Figures \ref{fig:varmB3}, \ref{fig:varmB12} and \ref{fig:varmB24}),  3 regions can be seen 
(easier to see for the iodine part than for the xenon part): two large regions for half lower and upper parts of the core and a small 
one for the central part of the core. The variances are lower for the median part than for the two others
 (also easier to see for the iodine than for the xenon)
\item  The variances are lower in the multivariate modelling than in the univariate modelling, that is to say  $tr(\mathbf{B}_{multi})  < tr(\mathbf{B}_u)$. But the multivariate modelling takes into account correlations between species: thus error statistics are spread all around the elements of the 
covariance matrix.
\end{itemize}

\begin{figure}[htbp]
\begin{center}
\subfloat[$\mathbf{B}_{d}$]{\label{fig:corrBd}}{\includegraphics[width=0.5\columnwidth]{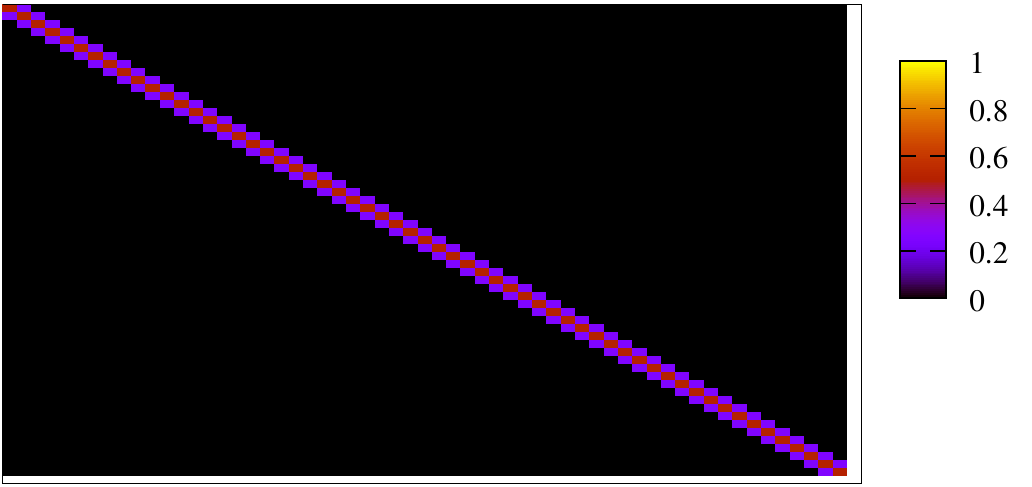}}\subfloat[$\mathbf{B}_{u}$]{\label{fig:corrBu}}{\includegraphics[width=0.5\columnwidth]{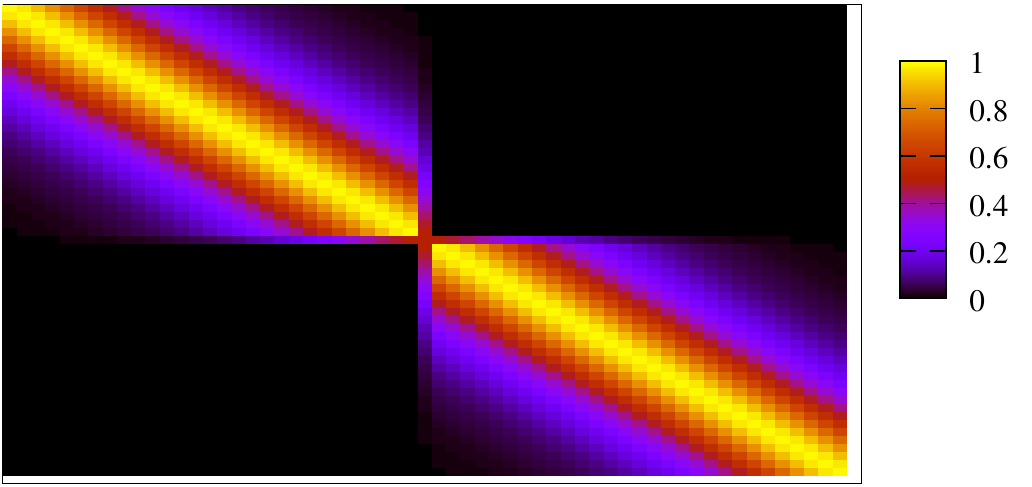}}
\subfloat[$\mathbf{B}_{3}$]{\label{fig:corrB3}}{\includegraphics[width=0.5\columnwidth]{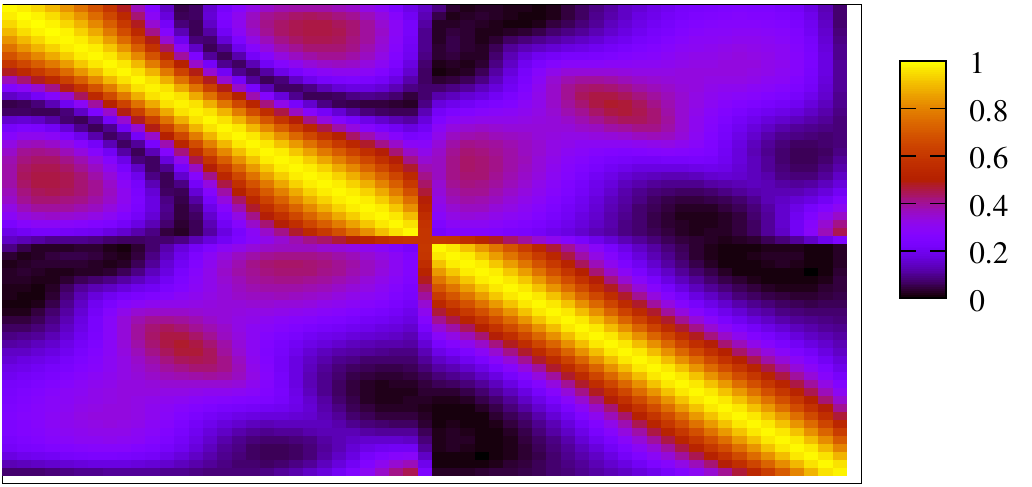}}\subfloat[$\mathbf{B}_{12}$]{\label{fig:corrB12}}{\includegraphics[width=0.5\columnwidth]{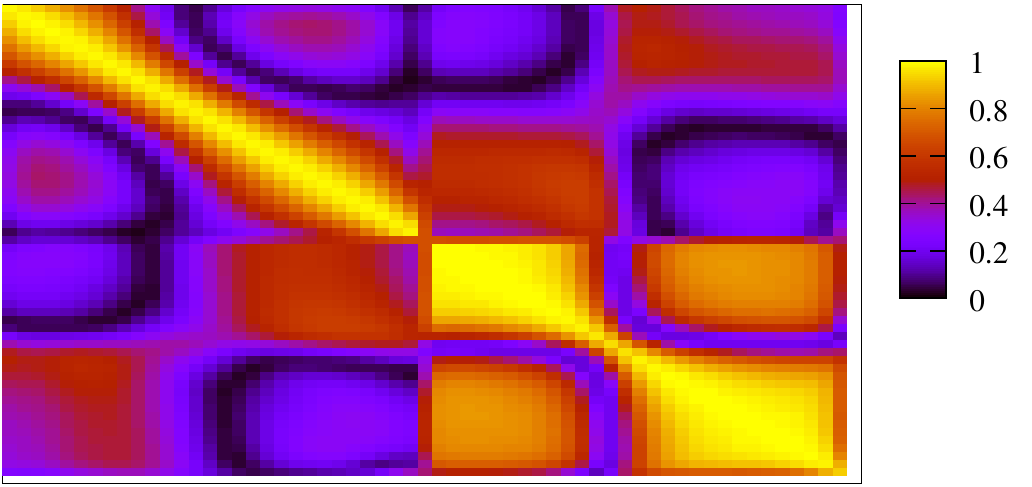}}
\subfloat[$\mathbf{B}_{24}$]{\label{fig:corrB24}}{\includegraphics[width=0.5\columnwidth]{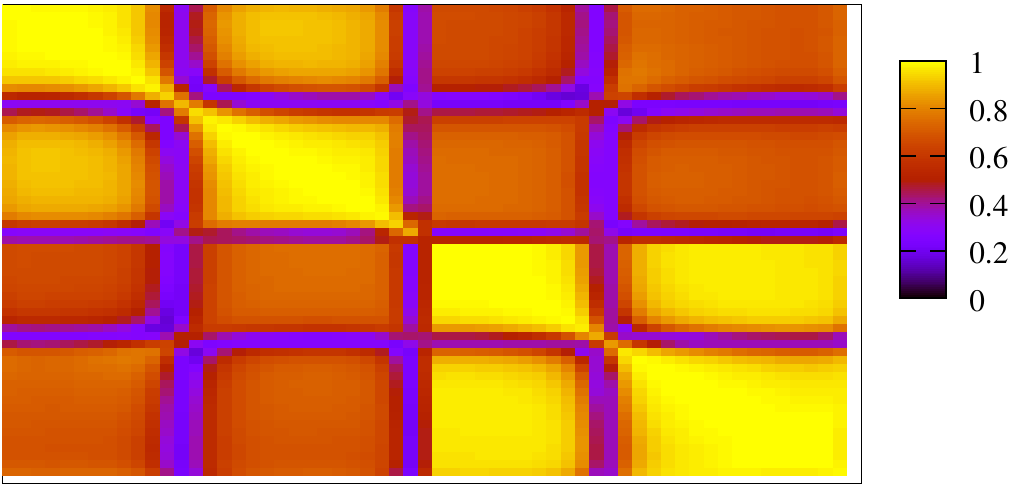}}
\end{center}
\caption{Absolute values of the components of the correlations related to the matrices $\mathbf{B}$ for univariate and multivariate modelling with respect to the nodes of the spatial mesh.}
\label{fig:corrB}
\end{figure}

To get an idea of the diffusion of the xenon error to the iodine error in the multivariate modelling,  Figure \ref{fig:corrB} gives the absolute values (for the clarity of the figure) of the correlations with 
respect to the nodes of the spatial mesh, where the correlations are defined as
$$
corr(i,j)=\mathbf{B}(i,j)/\sqrt{\mathbf{B}(i,i)\mathbf{B}(j,j)},
$$
where the element $(i,i)$ of the matrix $\mathbf{B}$ corresponds to the
variance of the background error on xenon at the node number $i$ (resp. iodine at the node number $i-30$) if $i<30$  (resp. $i>31$) and the  element $(i,j)$ (or $(j,i)$ since $\mathbf{B}$ is symmetric) corresponds to the
covariance of the background error between xenon and iodine at the node $i$.  
Then the range of the scale varies from 0 to 1.

We may notice:
\begin{itemize}
\item Compared to the univariate modellings for which  extra block diagonal terms are obviously null, 
introduction of spatial correlations fills in the two diagonal blocks in Figure \ref{fig:corrBu}. 
The choice of the correlation scale will be discussed later.
\item In the multivariate modellings (Figures \ref{fig:corrB3}, \ref{fig:corrB12} and \ref{fig:corrB24}), the correlation matrices present an internal
structure more or less pronounced with respect to the use of an evolution model on a more or less long time range.
As for the variances, one can see two large regions for the half upper and lower parts of the core and a narrower region for the central part 
of the core.
\begin{itemize}
\item Diagonal blocks: they  correspond  to the spatial correlations for a given species. Spatial correlations 
increase with the length of the time range used in the evolution model, first for the iodine correlations
 (e.g.  $\mathcal{M}_{12}$) and then for the xenon ones. This can be explained by the way of production of xenon  which 
is produced by radioactive decay of the iodine.
For the matrix $\mathbf{B}_{24}$,  correlations between upper and lower regions of the core are almost as strong as the spatial correlations inside these 
regions.
\item Extra-diagonal blocks: they  correspond to the correlations between the xenon and the iodine. 
The same correlation increase is noticed with the length of the time range used in the evolution model. 
But these correlations stay below the spatial correlation level inside the same species.
\item Central blocks: a region including very few nodes of the middle of the core seems to be insensitive to the time range used in the evolution model:
the background error for these nodes is slightly correlated to the background error of the other nodes. 
At last, one can notice that the correlation scale for these nodes seems to be shorter to the scale 4 set up in the correlation scale modelling described 
in Section \ref{sec:um}.
\end{itemize}
\end{itemize}

\section{COMPARISON BETWEEN 4DVAR AND 3DVAR SCHEMES}\label{sec:res}

We are in the twin experiment framework: the true state $\mathbf{X}^t$ is a simulated state. 
The simulated case was briefly described in the previous section.
We always assume that background and measurement errors are Gaussian. 
Errors introduced in measurements are set to 10\%, 5\% and 1\%
respectively for axial power, power axial offset and boron concentration measurements.

The purpose is to compare the result of the 3DVAR and 4DVAR schemes  at the assimilation time but also after at a forecast of 10 hour (typical time for xenon oscillation). 
In the case of the 3DVAR assimilation, various modelling of
the $\mathbf{B}$ matrix will be under consideration.  The 3DVAR results are compared to a  4DVAR result whose characteristics are set up according to~\cite{Poncot09}:  
 the  window size  is set to 6 hours and the observation frequency is set to 2 hours  (then $3$
observation sets are used)

The various $\mathbf{B}$ matrices used for the 3DVAR schemes are the following:
\begin{itemize}
\item $\mathbf{B}_d$ corresponds to the univariate modelling of $\mathbf{B}$ without any spatial correlation;
\item $\mathbf{B}_u$ corresponds to the univariate modelling with spatial correlation;
\item $\mathbf{B}_i$ corresponds to the multivariate modelling where the time range used in the evolution model equals to {\itshape i} hours (here 3, 12 or 24 hours).
\end{itemize}

Results are organized as follows: firstly we are analysing the 4DVAR results. Secondly we are comparing 1D errors on axial shape of xenon, iodine and power. 
Then we  show the time evolution of the mean error.
 At last we discuss  the statistics of the analysis given by the analysis matrices.

\subsection{4D VAR results}

\begin{figure}[htbp]
\begin{center}
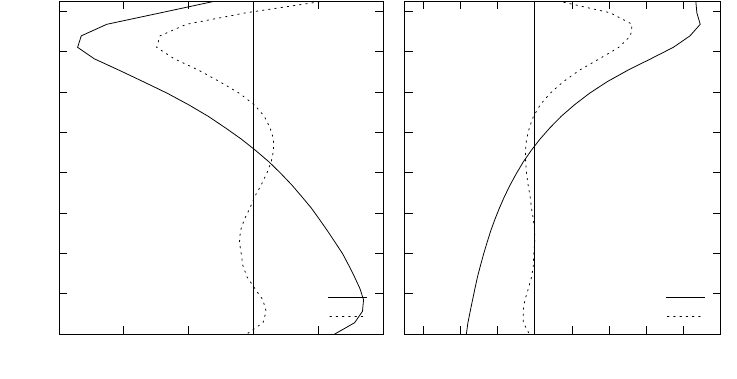
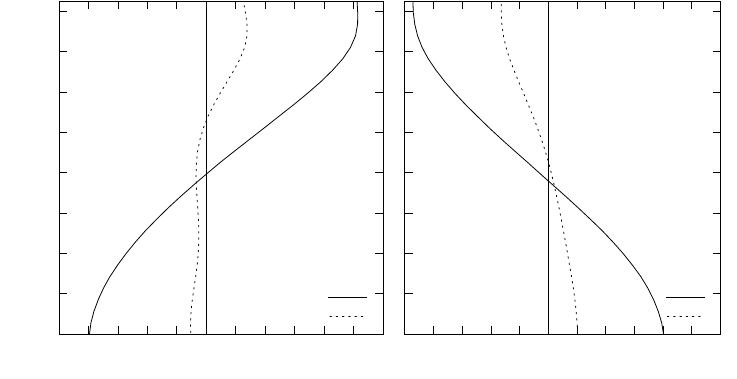
\caption{Relative errors on xenon, iodine at $t_0$ and on power at $t_0$ and $t_0+10h$ for the background state and the 4DVAR analysis state.}
\label{fig:bm}
\end{center}
\end{figure}

First we will look at the results of the 4DVAR scheme that is known to be the most efficient for
forecasting. With this  scheme, the modelling of xenon/iodine correlation for $\mathbf{B}$ 
has a weaker influence on the quality of the analysis  and then we use the $\mathbf{B}_u$ matrix.
Figure \ref{fig:bm} shows relative  errors\footnote{The ``true'' value of the fields xenon, iodine
and power is known since we are working in the framework  of twin experiments.} on  xenon, iodine
at the assimilation time $t_0$ and on power at $t_0$ and $t_0+10h$ for two computations:  the
background computation which gives the background state and a 4DVAR computation with the
assimilation characteristics given before.


 On this Figure \ref{fig:bm}, one can see that the 4DVAR computation allows to
reduce errors on xenon and iodine axial shape at least by a factor of 2. For the axial power shape
the decrease is even more important since the errors are reduced by a factor of 4. What follows
aims at showing that it is possible to approach the 4DVAR results quality with 3DVAR schemes using
multivariate  $\mathbf{B}$ matrices.

\subsection{Comparison between 4DVAR and 3DVAR }\label{subsec:4d3d}

\begin{figure}[htbp]
\begin{center}
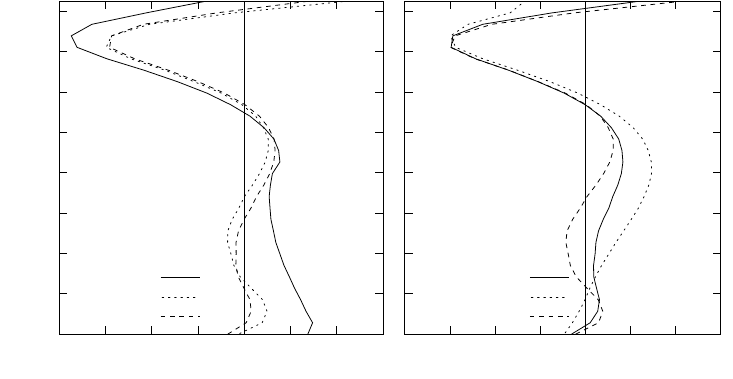
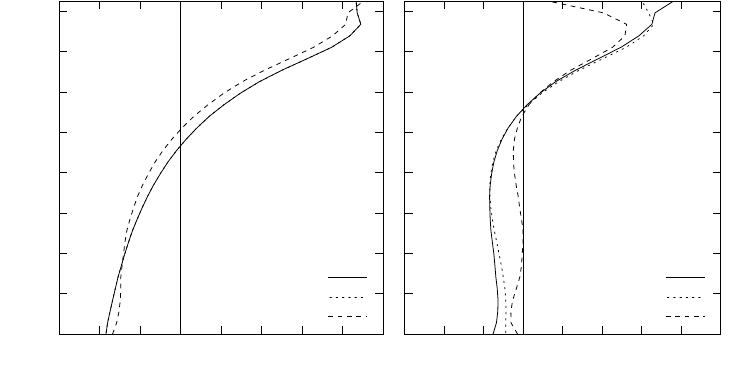
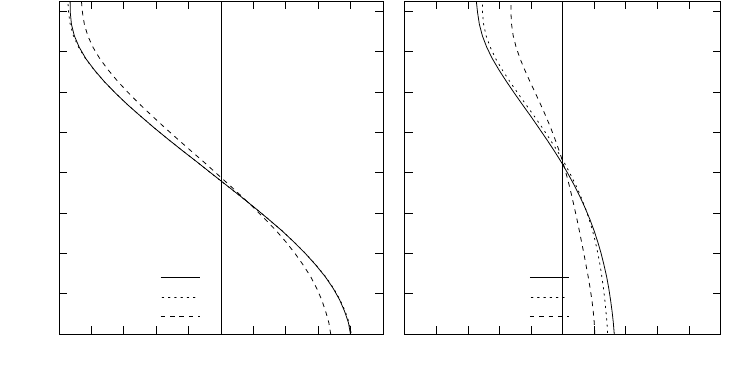
\caption{Relative errors on xenon, iodine at $t_0$ and on power at $t_0+10h$.
  At $t_0$ 3DVAR iodine analysis and background are mistaken.
A $t_0+10h$, 3DVAR and background predicted axial power shapes are mistaken.}
\label{fig:1Derror}
\end{center}
\end{figure}

Figure \ref{fig:1Derror} gives the 1D relative errors respect to the true state on axial shape of
xenon, iodine and power for the different  assimilation schemes. 
We see that the 4DVAR results are better than the 3DVAR ones 
independently of the choice of $\mathbf{B}$.
 However, looking at 3DVAR results at analysis time, we notice that xenon is rather well estimated
in all variational schemes except for the scheme which uses the elementary modelling of
$\mathbf{B}$. The xenon estimation is not as good for the bottom half as for the top half, but it
is everywhere  better than the state given by the background computation. This good result can be
explained by the fact that the xenon level at the assimilation time is directly related to the
assimilated observations, that is to say to the integrated powers. 

On the opposite, iodine is not directly related to the power level but through the production of
xenon since xenon is essentially produced by the radioactive decay of the iodine. Thus the
observation operator $\mathcal{H}$ does not depend on iodine. Therefore it is not possible to
correct iodine state with a 3DVAR scheme unless correlations between xenon and iodine are
introduced in $\mathbf{B}$. As a consequence the 3DVAR analysis error is equal to the background
one for the computation with the matrices $\mathbf{B}_d$ and $\mathbf{B}_u$. With a multivariate
modelling of $\mathbf{B}$ it can be seen that the time range used  to build the extra-species
correlations influences the  quality of the iodine analysis. It seems that the longer the time
range is, the better the analysis is. In fact there is no convergence towards the 4DVAR analysis
quality, and taking a 48 hours time range does not allow to improve 3DVAR results. One can assume
that the optimal time range is related to the time constants of radioactivity decay of the xenon
and iodine.

The knowledge of the iodine level is not important for the monitoring operator system. We are
expecting to improve axial power forecasts with the assimilation techniques. If we take a look at
the power estimation after 10 hours,  we notice that background and diagonal and univariate 3DVAR
analysis axial power shapes are very close.  Since xenon is essentially produced by radioactive
decay of iodine, a bad estimation of the initial iodine concentration will affect the  xenon
concentration estimation later.  As expected the multivariate modelling  leads to a significant
improvement in the 3DVAR forecast (up to a factor of 2 to 3 on the errors for the use of the
matrices $\mathbf{B}_{12}$ and $\mathbf{B}_{24}$). The multivariate modelling resulting from the
use of a time range of 3 hours in the evolution model does not seem to be  sufficient to really
improve power forecasts. As a first conclusion, the optimal time range seems to be around 12
hours: the 24 hours time range does not improve  much the forecasts compared to  the 12 hours time
range but it increases the amount of computation for $\mathbf{B}$.

\subsection{Time evolution of the $L^2$-norm error}

\begin{figure}[htbp]
\begin{center}
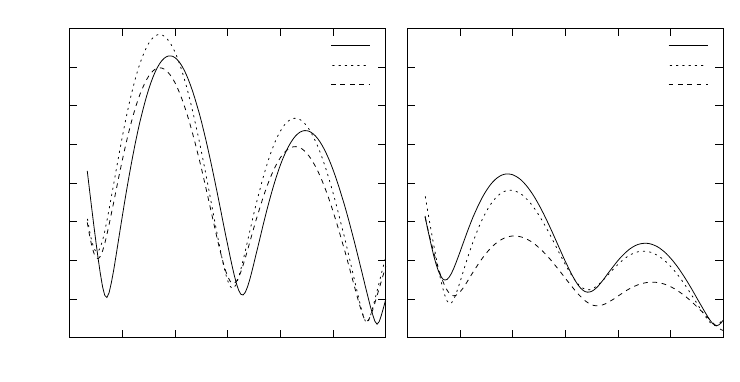
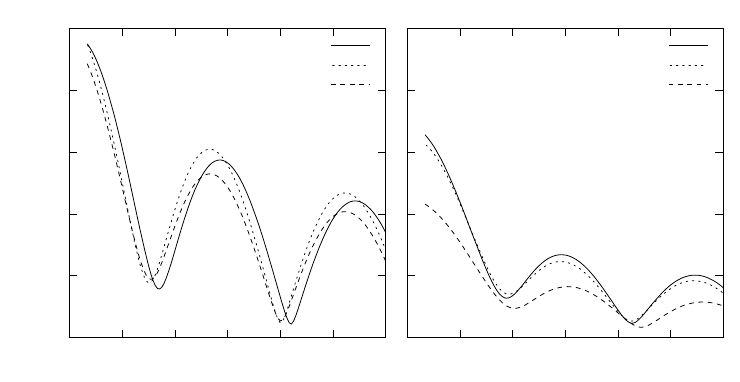
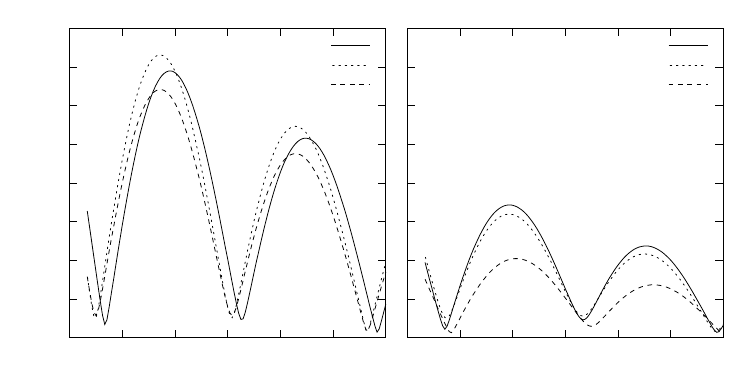
\caption{Relative $L^2$-norm errors on monodimensional xenon and iodine fields with respect to time for 3DVAR and 4DVAR schemes 
based on various background matrices.}
\label{fig:norm}
\end{center}
\end{figure}

We would like to confirm what has been shown at the assimilation time $t_0$ and $t_0+10$ hours.
Figure \ref{fig:norm} presents the time evolution of the $L^2$-norm error of the three studied
fields for the different matrices $\mathbf{B}$ used. An oscillation which tends to damp can be
seen. But it does not change the previously  drawn conclusion. Two groups can be seen: the first
one composed by the analysis computed  by the 3DVAR schemes using the matrices $\mathbf{B}_d$,
$\mathbf{B}_u$, $\mathbf{B}_3$; the second one composed by the analysis computed  by the 3DVAR
schemes using the multivariate  matrices $\mathbf{B}_{12}$, $\mathbf{B}_{24}$ plus the analysis
computed by the 4DVAR scheme based on the matrix $\mathbf{B}_u$. As one can see on the Figure
\ref{fig:norm}, the latter still represents a ''reference'' towards one can except to tend with a 
3DVAR scheme based on a multivariate  modelling. And the 3DVAR schemes where no or very few
correlation between  xenon and iodine is modelled are unable to give good forecasts.
%
We can deduce that the 3DVAR assimilation results
depending a lot of the $\mathbf{B}$ involved .


\subsection{Analysis matrices}

Assimilation techniques offer  {\itshape a posteriori} diagnostic on the computed
analysis by the mean of the analysis matrix.  
We use this opportunity to go more deeply into the study of the  respective quality of the various approaches.

\begin{figure}[htbp]
\begin{center}
\subfloat[$\mathbf{A}_{d}$]{\label{fig:varAd}{\includegraphics[width=0.5\columnwidth]{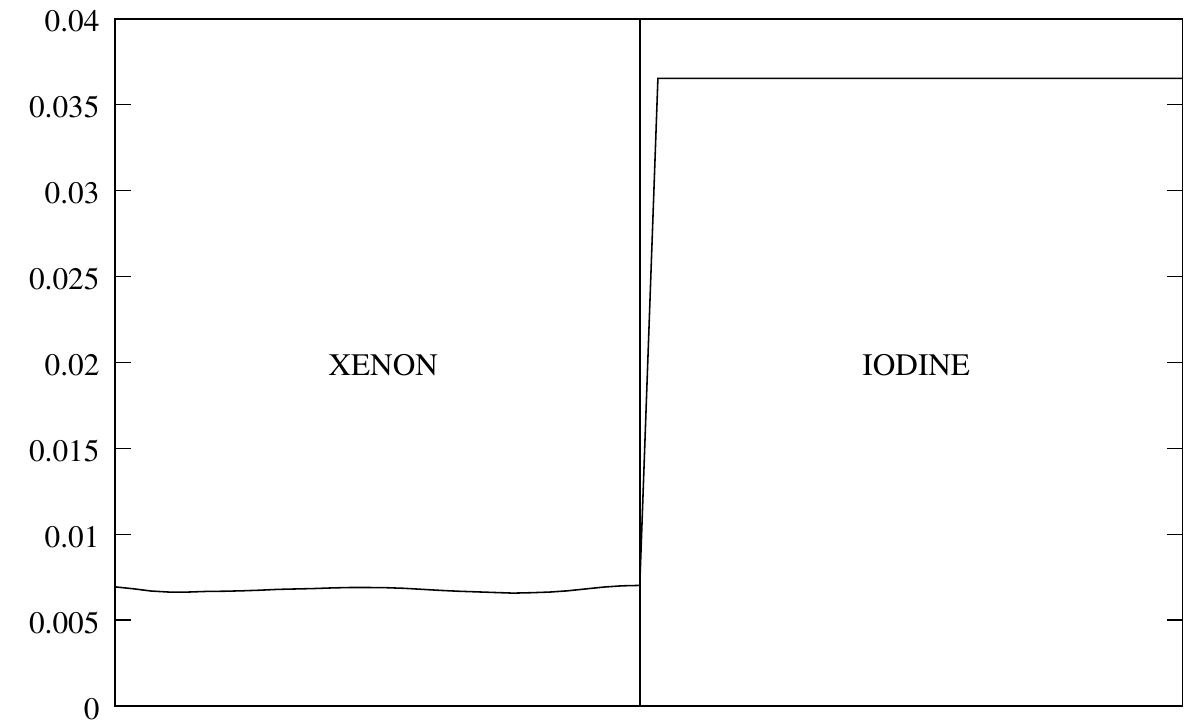}}}\subfloat[$\mathbf{A}_{u}$]{\label{fig:varAu}{\includegraphics[width=0.5\columnwidth]{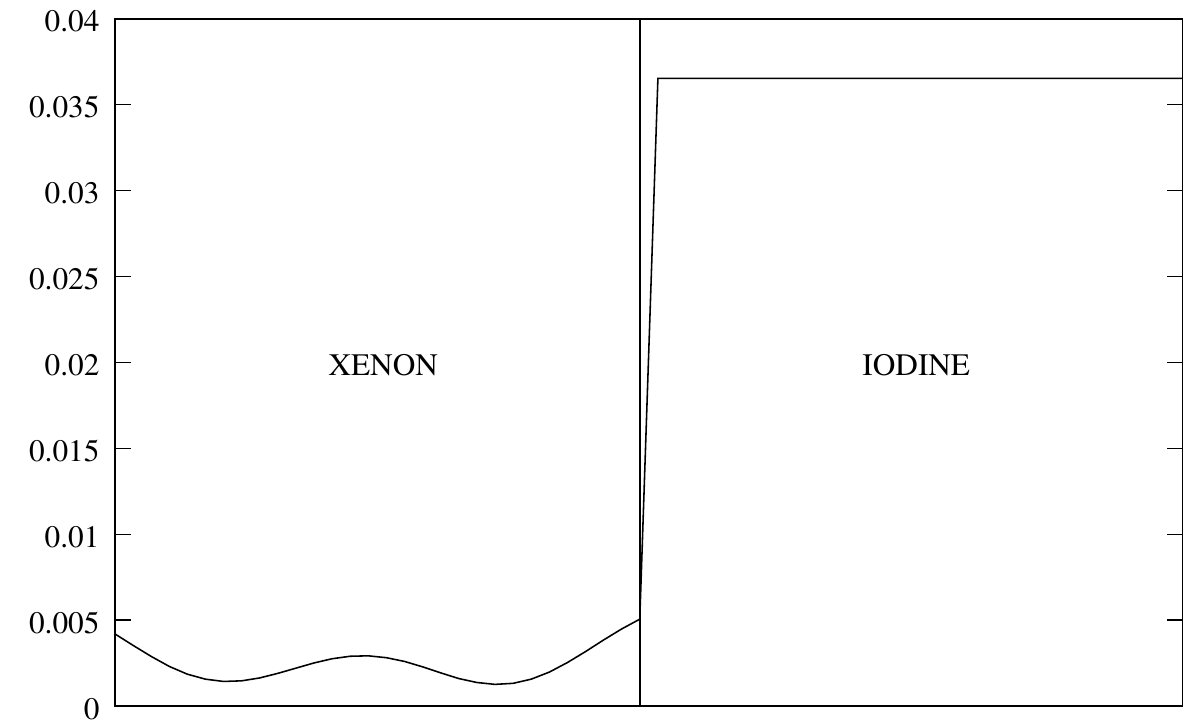}}}\\
\subfloat[$\mathbf{A}_{3}$]{\label{fig:varA3}{\includegraphics[width=0.5\columnwidth]{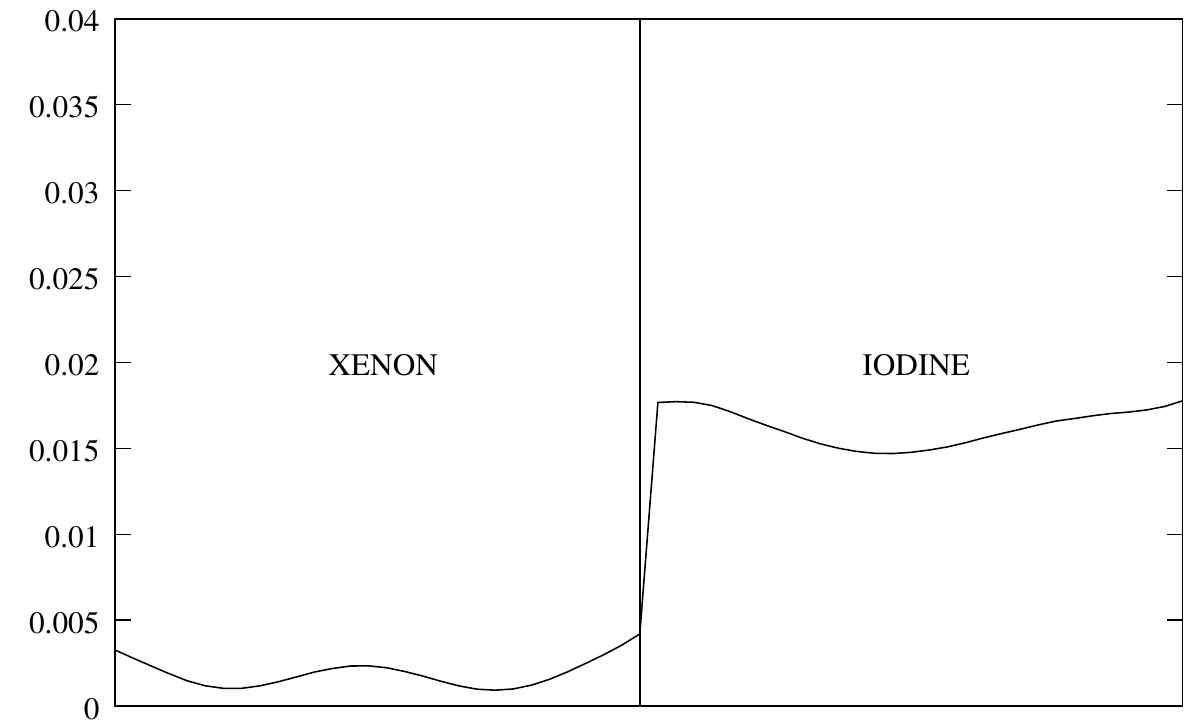}}}\subfloat[$\mathbf{A}_{12}$]{\includegraphics[width=0.5\columnwidth]{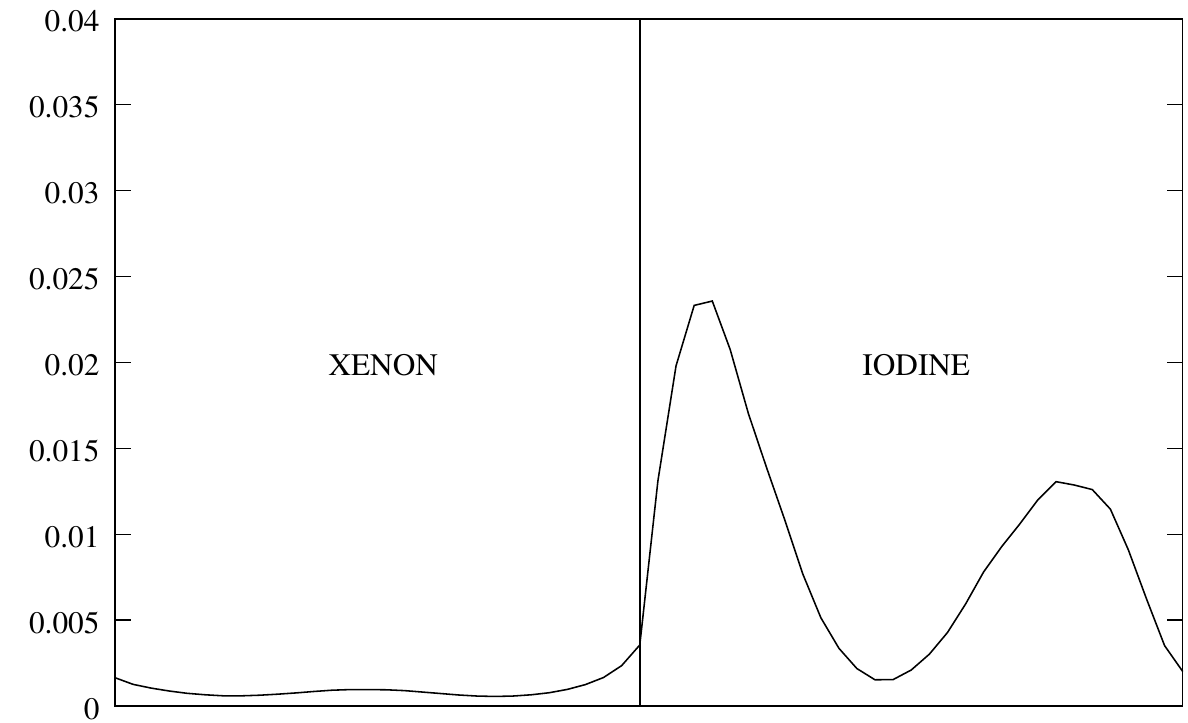}}\\
\subfloat[$\mathbf{A}_{24}$]{\includegraphics[width=0.5\columnwidth]{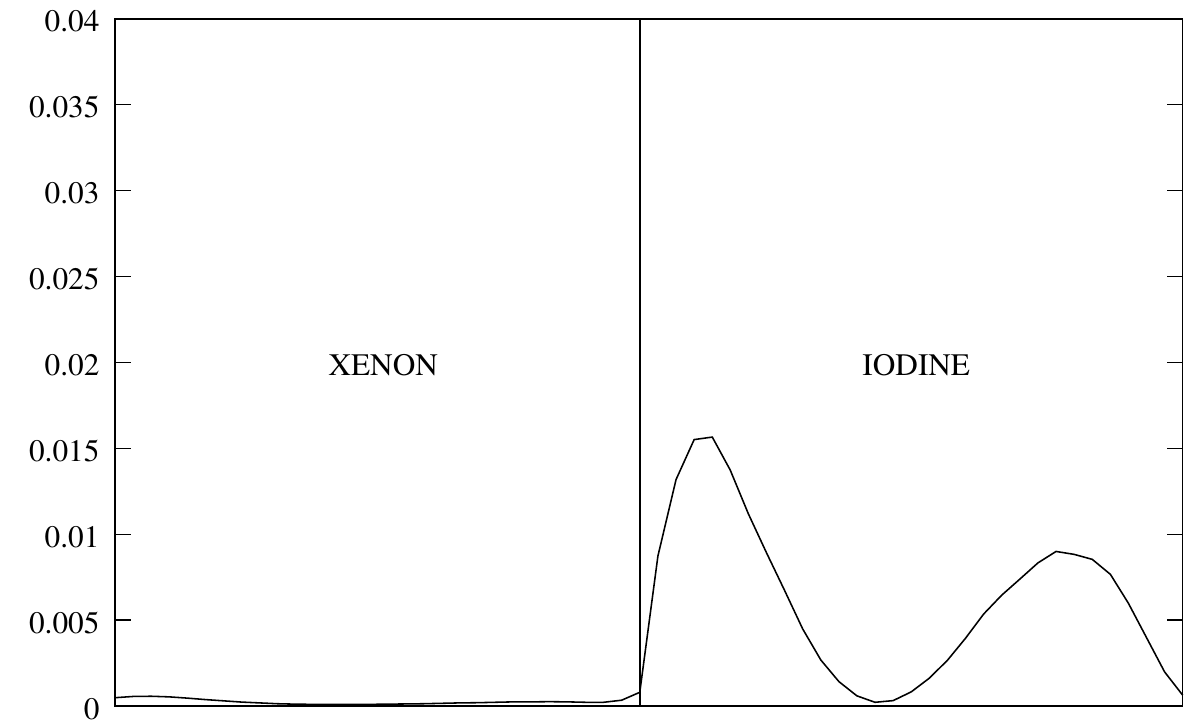}}\subfloat[$\mathbf{A}_{4DVAR}$]{\includegraphics[width=0.5\columnwidth]{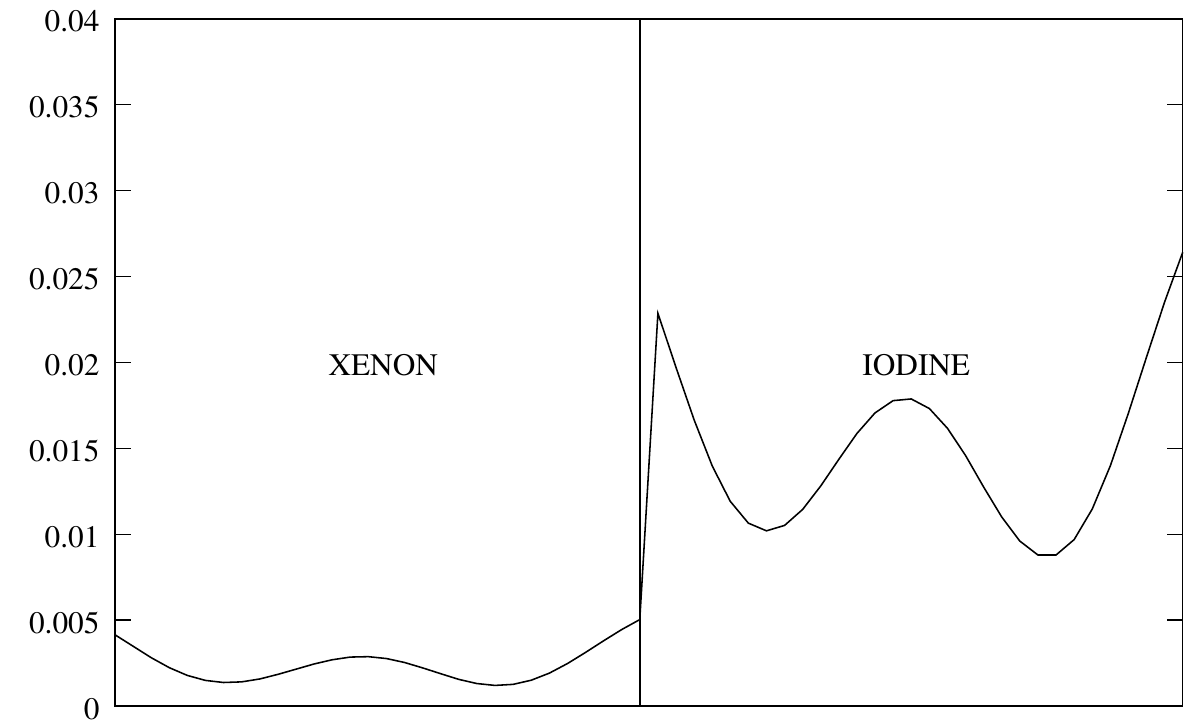}}
\end{center}
\caption{Diagonal terms of the analysis matrices $\mathbf{A}$ for 3DVAR and 4DVAR schemes based on univariate and multivariate
modelling of $\mathbf{B}$.}
\label{fig:varA}
\end{figure}

\begin{figure}[htbp]
\begin{center}
\subfloat[$\mathbf{A}_{d}$]{\label{fig:corrAd}{\includegraphics[width=0.5\columnwidth]{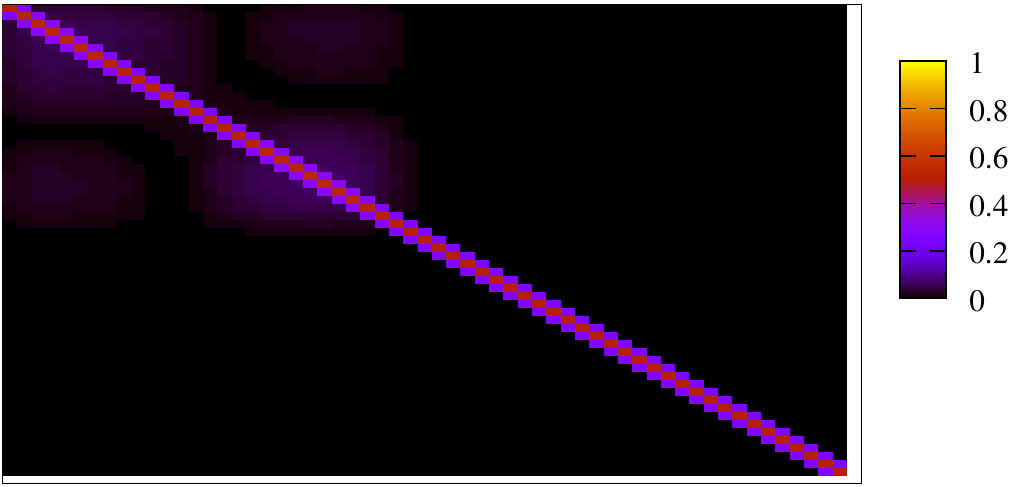}}}\subfloat[$\mathbf{A}_{u}$]{\label{fig:corrAu}{\includegraphics[width=0.5\columnwidth]{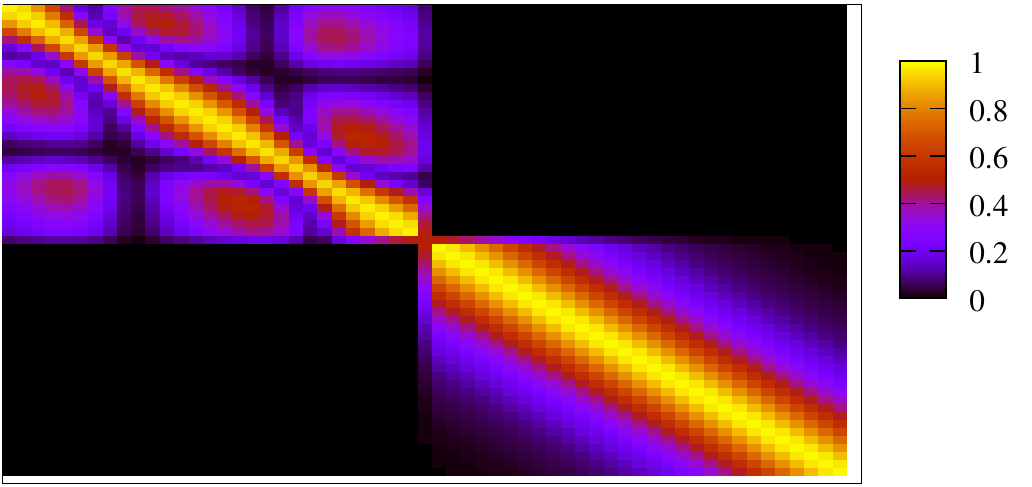}}}\\
\subfloat[$\mathbf{A}_{3}$]{\label{fig:corrA3}{\includegraphics[width=0.5\columnwidth]{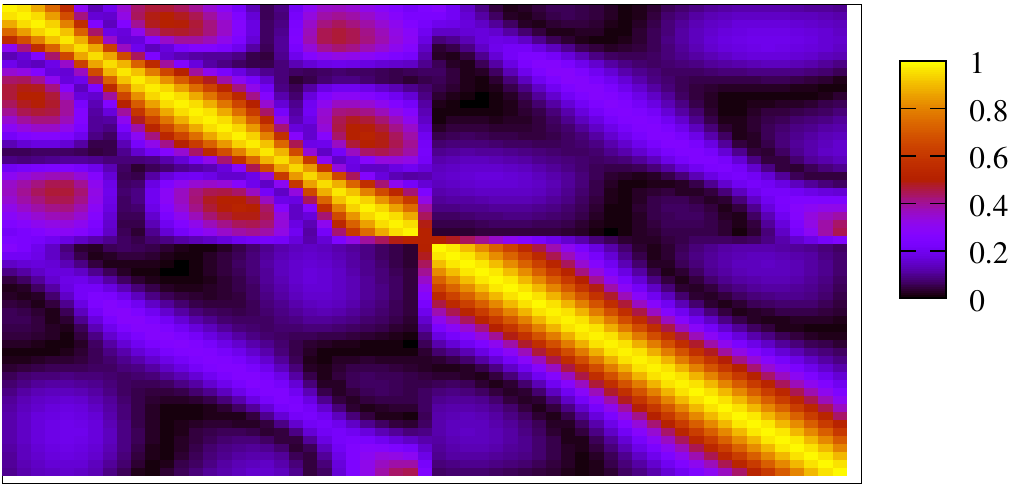}}}\subfloat[$\mathbf{A}_{12}$]{\label{fig:corrA12}{\includegraphics[width=0.5\columnwidth]{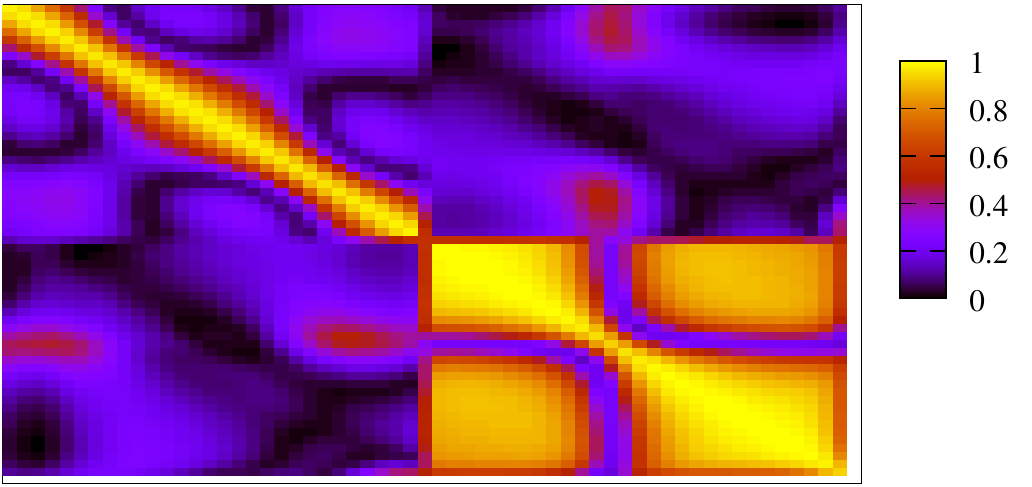}}}\\
\subfloat[$\mathbf{A}_{24}$]{\includegraphics[width=0.5\columnwidth]{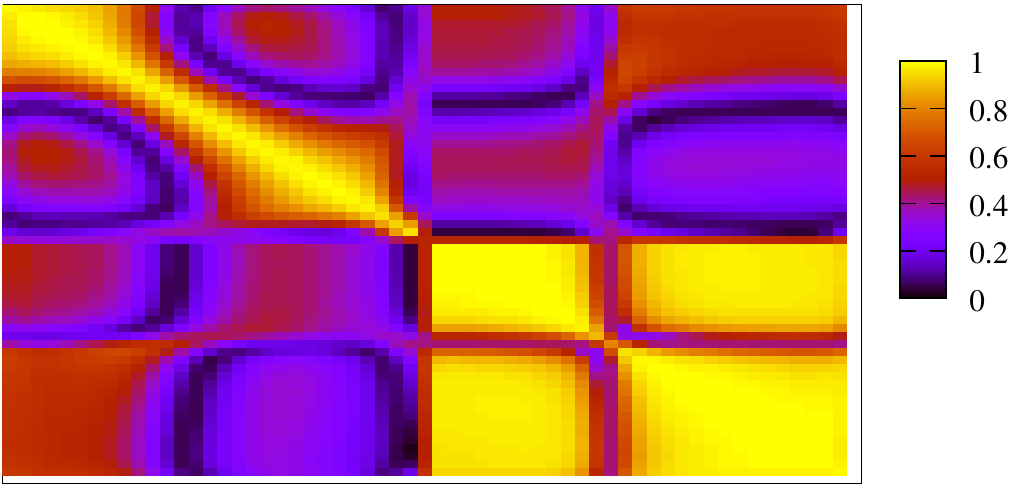}}\subfloat[$\mathbf{A}_{4DVAR}$]{\includegraphics[width=0.5\columnwidth]{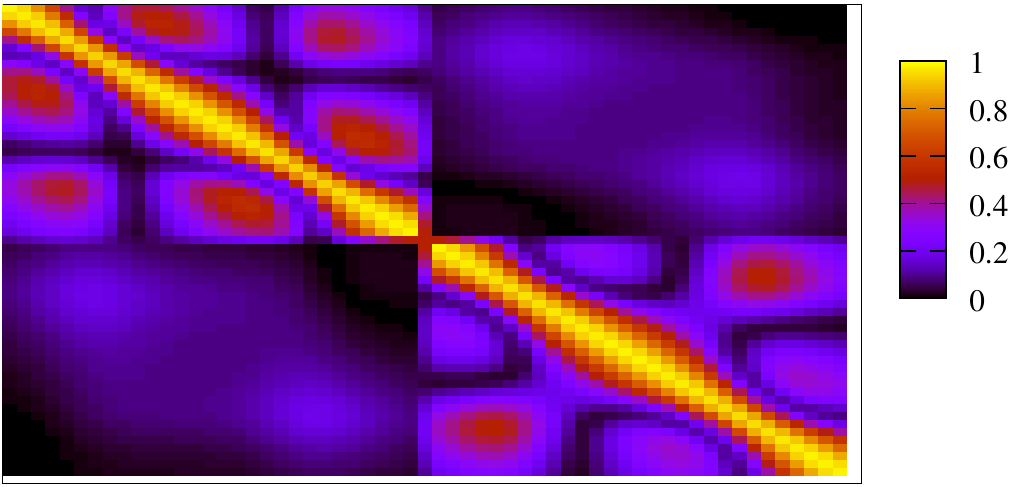}}
\end{center}
\caption{Absolute values of the components of the correlations related to the matrices $\mathbf{A}$ for 3DVAR and 4DVAR schemes based on univariate and multivariate modelling of $\mathbf{B}$.}
\label{fig:corrA}
\end{figure}

Figures \ref{fig:varA} and
\ref{fig:corrA} are to be compared to  Figures \ref{fig:varB} and
\ref{fig:corrB} that show the structure and amplitude of the elements of the
background matrix $\mathbf{B}$. Figures \ref{fig:varAd} and \ref{fig:varAu} show
that the 3DVAR schemes based on univariate process is not able to reduce the
variance (represented by the diagonal of $\mathbf{A}$) of the iodine error.
Though the scheme based on the multivariate matrix $\mathbf{A}_3$ has been
proved not being good for forecast, Figure \ref{fig:varA3} shows that it is much
better than the previous ones since it allows to reduce by a factor of 2 the
error on iodine.


Figure \ref{fig:corrA} gives absolute values of the correlations with respect to the nodes of the
spatial meshes. 
It is to be noted that if the matrix $\mathbf{B}$ does not contain correlations between the xenon
 and iodine errors (extra diagonal blocks), the
analysis matrix $\mathbf{A}$ issued from a  3DVAR scheme does not contain any correlation between them neither. The 4DVAR scheme is based on the univariate $\mathbf{B}_u$ but  the corresponding analysis matrix $\mathbf{A}_{4DVAR}$ shows correlations between xenon and iodine. These latter are brought by 
the 4DVAR scheme through the use of the evolution model on a 6 hours time range but they are shorter than the ones  in the matrices
$\mathbf{A}_3$, $\mathbf{A}_{12}$ and $\mathbf{A}_{24}$.

Figure \ref{fig:corrA} also presents the spatial correlations intra species (diagonal blocks). It
can be compared to the spatial correlations shown in Figure \ref{fig:corrB}. It can be seen that
spatial correlations after the assimilation process are a little bit shorter than they were in the
matrix $\mathbf{B}$.




\section{CONCLUSIONS}

In this paper, we have shown how variational data assimilation methods can be used to improve the
accuracy of the prediction of the xenon concentration in PWRs. A monodimensional xenon dynamics code CIREP1D was developed for this purpose.

The investigation done here in twin experiments proves that the 4DVAR scheme is a very efficient method 
 to improve the accuracy of the prediction of the xenon concentration as
well as the axial power shape in PWRs.  
However this method is computationally  costly
and the development  of the adjoint of the model is mandatory.

A computationally cheaper solution is the 3DVAR that can lead to rather good
result. Nevertheless these latters can only be obtained through a careful
modelling on the associated  background error covariance matrix
$\mathbf{B}$. Among  the various modellings of  $\mathbf{B}$ studied here the one
based on the multivariate modelling  is the most satisfactory.  

The next stage  for the 3DVAR approach could consist in setting up an assimilation chain where the
$\mathbf{B}$ matrix is updated at each analysis step as it is successfully done
for meteorological operational forecasts.

\bibliographystyle{elsarticle-num}
\bibliography{biblio}

\newcommand\SetFigures{\clearpage\vspace*{4.0cm}\begin{center} \Huge Figures
\normalsize\end{center}\clearpage}

\clearpage

\listoffigures
\clearpage


\end{document}